\documentclass[]{svjour3}
\smartqed

\makeatletter
\def\switch@array{}
\makeatother

\usepackage[T1]{fontenc}
\usepackage[utf8]{inputenc}
\usepackage{lmodern}
\usepackage{amsmath,amssymb,mathtools}
\usepackage{graphicx}
\usepackage{booktabs}
\usepackage{longtable}
\usepackage{array}
\usepackage{multirow}
\usepackage{xcolor}
\usepackage{enumitem}
\usepackage{tikz}
\usetikzlibrary{decorations.pathmorphing}
\usepackage{tabularx}
\usepackage{cite}
\usepackage{hyperref}
\hypersetup{colorlinks=true,linkcolor=blue,citecolor=blue,urlcolor=blue}

\journalname{European Physical Journal C}

\usepackage{geometry}
\newcommand{\be}{\begin{equation}}
\newcommand{\ee}{\end{equation}}
\def\bea{\begin{eqnarray}}
\def\eea{\end{eqnarray}}

\begin{document}

\title{Forward backward CP Asymmetry in $\tau^-\to K\pi\nu_\tau$ in the Left–Right Inverse Seesaw Model}

\author{David Delepine \and Shaaban Khalil }

\institute{
David Delepine \at Division de Ciencias e Ingenier\'ias,  Universidad de Guanajuato, C.P. 37150, Le\'on, Guanajuato, M\'exico. \email{delepine@ugto.mx} \and
Shaaban Khalil \at Centre for Theoretical Physics, Zewail City of Science and Technology, 6th October City, 12588, Giza, Egypt.\email{skhalil@zewailcity.edu.eg}}
\date{Received: date / Accepted: date}
\maketitle

\begin{abstract}
The Left--Right Inverse Seesaw (LRIS) model, in which TeV-scale right-handed neutrinos can carry $\mathcal{O}(1)$ Yukawa couplings while light neutrino masses remain protected by a small lepton-number-violating scale, provides a testable link between the neutrino-mass mechanism and flavor and collider observables: the same non-decoupling dynamics has recently been shown to enable viable TeV-scale non-thermal leptogenesis and to explain the $B\to K\mu^+\mu^-$ anomaly. Here we examine a further, independent test of the same scalar sector in the semileptonic decay $\tau \to K\pi\nu_\tau$.  We identify a distinct, unsuppressed signal in the \emph{differential} forward-backward CP asymmetry $A_{\rm CP}^{\rm FB}(s)$, driven by interference between the SM vector current and a non-decoupling scalar operator $g_S$ generated by a top-quark flavor-changing neutral-current box diagram with internal heavy neutrinos and charged Goldstone bosons. We derive the effective $|\Delta S|=1$ Hamiltonian, verify consistency with $K^0$--$\bar K^0$ and $B_d^0$--$\bar B_d^0$ mixing, $B\to X_s\gamma$, and neutrino non-unitarity constraints, and show numerically that $A_{\rm CP}^{\rm FB}(s)$ is resonantly enhanced near the $K_0^*(1430)$ state, reaching $\mathcal{O}(10^{-4})$ -- within reach of Belle~II.
\end{abstract}


\section{Introduction}
\label{sec:intro}

TeV-scale extensions of the Standard Model (SM) gauge group remain one of the most compelling avenues for addressing outstanding puzzles that the SM leaves unanswered, from the origin of neutrino masses to the matter-antimatter asymmetry of the Universe. Among these, the Left-Right symmetric extension $SU(3)_C \times SU(2)_L \times SU(2)_R \times U(1)_{B-L}$~\cite{Pati:1974yy,Mohapatra:1974hk,Senjanovic:1975rk} is particularly attractive: it restores parity as a fundamental symmetry spontaneously broken at the $v_R$ scale, and naturally accommodates small neutrino masses once combined with a seesaw mechanism. In this work we focus on its \emph{inverse-seesaw} realization (LRIS)~\cite{Khalil:2007dr,Dev:2012sg,Barry:2013xxa}, which departs from the canonical type-I seesaw in a phenomenologically important way: light neutrino masses are protected by a small lepton-number-violating scale $\mu_S$ rather than by parametrically suppressing the neutrino Yukawa couplings. As a result, the right-handed neutrino sector can carry $\mathcal{O}(1)$ Yukawa couplings even for TeV-scale masses $M_{N_i}$, opening the possibility of \emph{unsuppressed}, experimentally accessible signatures in flavor and collider observables -- a qualitative departure from the decoupled, largely inert heavy-neutrino sector of the canonical seesaw.

This TeV-scale accessibility is not merely a formal feature of the model. In a closely related $U(1)_{B-L}$ extension with an inverse-seesaw neutrino sector, we have recently shown that the same $\mathcal{O}(1)$ Yukawa structure that would render conventional \emph{thermal} leptogenesis ineffective (due to strong washout) instead enables a viable \emph{non-thermal} leptogenesis scenario operating entirely at the TeV scale, providing a direct and testable link between the neutrino mass mechanism and the cosmological baryon asymmetry~\cite{Delepine:2026lepto}. Within the LRIS model specifically, the same non-decoupling dynamics that we exploit below for $\tau \to K\pi\nu_\tau$ has also been shown to generate a viable explanation of the long-standing $B \to K\mu^+\mu^-$ anomaly: a charged-scalar/heavy-neutrino box diagram produces an unsuppressed contribution to the vector Wilson coefficient $\Delta C_9^\mu$ while a GIM-like phase structure in the right-handed quark mixing matrix keeps $B_s$--$\bar{B}_s$ mixing and $b \to s\gamma$ safely under control~\cite{Delepine:2026bkmumu}. Taken together, these results indicate that the LRIS model is a rich and internally consistent framework in which the same non-decoupling box-diagram mechanism, rooted in the inverse-seesaw structure, correlates observables as disparate as baryogenesis, rare $B$-meson decays, and -- as we show in this work -- semileptonic $\tau$ decays. Establishing such correlated, falsifiable predictions across otherwise unrelated sectors is, in our view, a more robust test of the model than any single anomaly.

With this broader motivation in mind, we turn to the semileptonic decay $\tau \to K\pi\nu_\tau$, which offers a complementary and largely unexplored window onto the LRIS scalar sector. Unlike purely leptonic decays, the presence of hadronic final-state interactions (FSI) generates the strong phases necessary to observe direct CP violation in the presence of new weak phases. In the channel $\tau^- \to \bar{K}^0 \pi^- \nu_\tau$, a SM CP violating asymmetry is expected due to $K^0$--$\bar{K}^0$ mixing~\cite{Grossman:2011zk}. Interest in this channel intensified following the BaBar collaboration's report of a $2.8\sigma$ discrepancy with the SM in the integrated CP asymmetry in the channel $\tau^- \to K_S^0 \pi^- \nu_\tau$~\cite{Lees:2012qi},
\begin{equation}
A_{\rm CP}^{\rm exp.} = -(0.36 \pm 0.23 \pm 0.11)\%\,.
\end{equation}
In the SM, the process is dominated by the charged-current interaction mediated by the $W$ boson. Focusing only on direct CP violation, the expected SM direct CP asymmetry is very small, $A_{\rm CP}^{\rm SM} \sim 10^{-12}$~\cite{Bigi:2005ts}, since a nonzero CP asymmetry requires both a weak phase and a strong rescattering phase. This makes $\tau\to K\pi\nu_\tau$ an attractive channel to search for Beyond the SM (BSM) scalar, tensor, or right-handed-current effects, and the BaBar measurement provides a useful, if not unique, point of reference for the sensitivity that current and near-future experiments can achieve in this channel.

While many NP models attempt to explain the BaBar anomaly through loop-induced tensor operators~\cite{Devi:2013gya,Cirigliano:2017tqn,Dhargyal:2016kwp}, such mechanisms typically suffer from a severe $1/16\pi^2$ suppression. In the context of TeV-scale physics, this leads to an integrated asymmetry of $\mathcal{O}(10^{-9})$, as has been shown in several studies~\cite{Cirigliano:2017tqn}. This suggests that the integrated asymmetry might serve better as a null test or an exclusion tool for loop-level tensor mediators rather than a primary discovery channel for extended gauge or scalar sectors -- a point that, as we show below, applies directly to the LRIS model itself.

CP violation (CPV) has been firmly established in non-leptonic decays of kaons~\cite{Christenson:1964fg}, as well as $B_d$~\cite{Aubert:2001nu}, $B_s$~\cite{Aaij:2013oba}, and more recently $D$ mesons~\cite{Aaij:2019kcg}. In the leptonic sector, CPV arises through neutrino mixing, which introduces a complex phase $\delta_{\rm CP}$ in the lepton mixing matrix~\cite{Fukuda:1998mi,Ahmad:2002jz}. This phase leads to CP-violating effects in neutrino oscillations and new generation of neutrino experiments are looking to measure it \cite{Abe:2019vii,Acero:2019ksn}.

The CPV observed in channels as $K_L^0 \to \pi^- \ell^+ \nu_\ell$~\cite{Christenson:1964fg} and $K_L^0 \to \pi^+\pi^- e^+ e^-$~\cite{Abouzaid:2006kk} originate from CP violation in the meson sector. Within the SM, direct CPV generated by leptonic sector is expected to be extremely suppressed.

Assuming the quark sector as the CPV origin, direct CP violation should be universal in channel decay modes  as $\tau^- \to \bar{K}^0 \pi^- \nu_\tau$ and $\tau^- \to K^- \pi^0 \nu_\tau$, since both are governed by the same underlying quark-level transition, $\tau^- \to s\bar{u}\nu_\tau$. Consequently, any genuine discrepancy between these channels would signal new physics. The possibility of direct, angular-integrated CPV in semihadronic tau decays, particularly in channels such as $\tau^- \to K_S \pi^- \nu_\tau$ and $\tau^- \to K^- \pi^0 \nu_\tau$, has been extensively studied in the literature~\cite{Bigi:2005ts,Calderon:2007rg,Devi:2013gya,Cirigliano:2017tqn,Dhargyal:2016kwp,Dhargyal:2016jgo,Kuhn:1996dv,Tsai:1996ps,Choi:1998yx,Delepine:2005tw,Delepine:2006fv,Delepine:2007qg,Kimura:2014wsa}.

In this work, we investigate $\tau \to K\pi\nu_\tau$ within the LRIS model introduced above, whose particle content includes a $9\times 9$ neutral fermion mass matrix and an extended scalar sector with a heavy charged Higgs $H^\pm$. We stress at the outset that, as we demonstrate explicitly in Sec.~\ref{sec:contrib}, the LRIS model does \emph{not} generate a sizable tensor operator and therefore \emph{cannot} account for the BaBar integrated-asymmetry anomaly: stringent phenomenological constraints on the right-handed quark mixing angles and scalar Yukawa couplings, arising from $K^0$--$\bar{K}^0$ mixing~\cite{Bona:2007vi}, $B_d^0$--$\bar{B}_d^0$ mixing, $D^0$--$\bar{D}^0$ mixing~\cite{Artuso:2015swg}, and precision electroweak tests~\cite{Senjanovic:2015yea}, combine with neutrino non-unitarity bounds from tau decays~\cite{Antusch:2014woa} to keep the predicted $A_{\rm CP}$ at or below $\mathcal{O}(10^{-7})$, several orders of magnitude below experimental sensitivity. This is, in fact, the generic conclusion for \emph{any} TeV-scale model in which the leading new-physics effect enters through a loop-suppressed tensor operator, as discussed above.

Rather than treating this as a negative result, we ask a logically independent question: does the LRIS scalar sector produce \emph{any} observable signature in this channel? We show that it does. Unlike the integrated asymmetry, the forward-backward CP asymmetry $A_{\rm CP}^{\rm FB}(s)$ is sensitive not to the (absent) tensor operator but to interference between the SM vector current and a genuinely different object: a scalar operator $g_S$ generated by a non-decoupling box diagram involving internal heavy neutrinos $N_i$, the top quark, and charged Goldstone bosons, in which the heavy-neutrino mass dependence cancels in the limit $M_{N_i} \gg v_R$~\cite{Pilaftsis:1991ug} -- the same non-decoupling mechanism responsible for the $B\to K\mu^+\mu^-$ result of Ref.~\cite{Delepine:2026bkmumu}. This allows $g_S$ to remain unsuppressed by heavy mass scales, leading to a predicted $A_{\rm CP}^{\rm FB}(s) \sim \mathcal{O}(10^{-4})$ that is within the potential reach of Belle~II~\cite{Kou:2018nap}. The differential forward-backward asymmetry thus provides an experimentally accessible, and falsifiable, probe of the LRIS scalar sector that is entirely independent of whether the model addresses the BaBar anomaly.

The remainder of this paper is organized as follows. In Sec.~II, we present the effective Hamiltonian for the four-fermion transition $\tau \to \nu_i s\bar{u}$, and define the CP observables in $\tau \to K\pi\nu_\tau$, with particular emphasis on the integrated CP asymmetry and the forward-backward CP asymmetry. In Sec.~III, we discuss the SM contributions to these observables. In Sec.~IV, we introduce the LRIS model, focusing on the relevant particle spectrum and Yukawa couplings for the process under consideration. Sec.~V is devoted to the LRIS contributions to CP asymmetry in $\tau \to K\pi\nu_\tau$, including the tree-level charged Higgs contribution, the flavor constraints on the model (including $B_d^0$--$\bar{B}_d^0$ mixing and the viable inverse-seesaw parameter space), and the box-diagram contribution involving internal heavy neutrinos $N_i$ and the top quark. Numerical results are presented in Sec.~VI, and our conclusions are given in Sec.~VII.

\section{Effective Framework and CP Observables in $\tau \to K\pi\nu_\tau$}
\label{sec:eff}
To describe the decay $\tau \to K\pi\nu_\tau$, heavy degrees of freedom are integrated out and the interaction is expressed in terms of an effective four-fermion Hamiltonian at the scale $\mu \sim m_\tau$. The most general $|\Delta S|=1$ effective Hamiltonian relevant for this decay can be written as~\cite{Gonzalez-Alonso:2016yxt,Miranda:2019sgo}
\begin{align}
\mathcal{H}_{\text{eff}}=
-\frac{G_F}{\sqrt{2}}V_{us}^\ast
\Big[
&(1+g_V)(\bar s\gamma_\mu u)(\bar\nu_i\gamma^\mu\tau)
-(1+g_A)(\bar s\gamma_\mu u)(\bar\nu_i\gamma^\mu\gamma_5\tau) +g_S(\bar s u)(\bar\nu_i\tau)
\nonumber\\
&+g_P(\bar s u)(\bar\nu_i\gamma_5\tau)
+g_T(\bar s\sigma_{\mu\nu}u)(\bar\nu_i\sigma^{\mu\nu}(1+\gamma_5)\tau)
\Big].
\label{eq:Heff}
\end{align}
Here $G_F$ denotes the Fermi constant and $V_{us}$ the relevant CKM matrix element. The coefficients $g_V$, $g_A$, $g_S$, $g_P$, and $g_T$ represent possible new physics contributions to the vector, axial-vector, scalar, pseudoscalar, and tensor operators, respectively. In the SM at tree level, $g_V = g_A = g_S = g_P = g_T = 0$. Contributions from right-handed vector currents are neglected since the corresponding gauge boson $W'$ is assumed to be very heavy, $M_{W'} \gg$ TeV~\cite{Senjanovic:2015yea}.

The hadronic side of the decay is encoded in three form factors, $F_+(s)$, $F_0(s)$, and $F_T(s)$, associated respectively with the vector, scalar, and tensor quark bilinears $\bar s\gamma_\mu u$, $\bar s u$, and $\bar s\sigma_{\mu\nu}u$~\cite{Jamin:2000wn,Bernard:2014vza,Gonzalez-Alonso:2016yxt}. $F_+(s)$ is dominated by the $K^*(892)$ resonance, with a subleading contribution from the excited $K^*(1410)$ state, while $F_0(s)$ is dominated by the scalar $K_0^*(1430)$ resonance together with a non-resonant $S$-wave background, conventionally described by the LASS parameterization~\cite{Aston:1987ir}. The explicit hadronic matrix elements, quark-mass inputs, and the resulting angular coefficients $\mathcal A(s)$, $\mathcal B(s)$, $\mathcal C(s)$ of the double-differential decay distribution are standard and are collected in Appendix~\ref{app:formfactors} for completeness; here we quote only the two results needed below. The double-differential decay distribution in $s=q^2$ and the helicity angle $\theta$ between the kaon momentum and the parent $\tau$ direction in the $K\pi$ rest frame is
\begin{equation}
\frac{d^2\Gamma}{ds\,d\cos\theta}
=
\frac{G_F^2|V_{us}|^2}{128\pi^3 m_\tau^3}
\left(1-\frac{s}{m_\tau^2}\right)^2
\frac{|\vec p_K|}{\sqrt{s}}
\left[
\mathcal A(s)
+
\mathcal B(s)\cos\theta
+
\mathcal C(s)\cos^2\theta
\right],
\label{eq:diffdist}
\end{equation}
and the angular coefficient controlling the forward-backward observables used throughout this work is
\begin{equation}
\mathcal B(s) =
\kappa_{VS}(s)\,
\text{Re}\!\left[(1+g_V)g_S^\ast\,F_+(s)F_0^\ast(s)\right]
+
\kappa_{VT}(s)\,
\text{Re}\!\left[(1+g_V)g_T^\ast\,F_+(s)F_T^\ast(s)\right],
\label{eq:Bcoeff}
\end{equation}
with kinematic weights
\begin{equation}
\kappa_{VS}(s)=-2\,\frac{m_\tau}{\sqrt{s}}\,|\vec p_K|,
\qquad
\kappa_{VT}(s)=\frac{16\,|\vec p_K|^2\sqrt{s}}{m_K+m_\pi}.
\label{eq:kappa}
\end{equation}
$\mathcal B(s)$ is CP-even, but its difference between $\tau^-$ and $\tau^+$ decays is CP-odd and proportional to the imaginary parts of the Wilson coefficients and form factors -- this is the quantity that ultimately drives $A_{\rm CP}^{\rm FB}(s)$ below.

\subsection{CP Asymmetry Observables}

For this process, one can define both the integrated CP asymmetry and the forward-backward CP asymmetry. The total integrated CP asymmetry is defined as~\cite{Bigi:2005ts,Grossman:2011zk}
\begin{equation}
A_{\rm CP}=
\frac{
\Gamma(\tau^- \to K^- \pi^0 \nu_\tau)
-
\Gamma(\tau^+ \to K^+ \pi^0 \bar\nu_\tau)
}{
\Gamma(\tau^- \to K^- \pi^0 \nu_\tau)
+
\Gamma(\tau^+ \to K^+ \pi^0 \bar\nu_\tau)
}
=
\frac{
\int_{s_{\rm min}}^{s_{\rm max}} ds\,
\left(
\frac{d\Gamma}{ds}
-
\frac{d\bar\Gamma}{ds}
\right)
}{
\int_{s_{\rm min}}^{s_{\rm max}} ds\,
\left(
\frac{d\Gamma}{ds}
+
\frac{d\bar\Gamma}{ds}
\right)
},
\label{eq:ACPdef}
\end{equation}
with $s_{\rm min}=(m_K+m_\pi)^2$ and $s_{\rm max}=m_\tau^2$.

A non-zero integrated CP asymmetry requires both a CP-violating weak phase and a non-trivial strong-phase difference between the interfering hadronic amplitudes. In the presence of tensor interactions, the dominant contribution typically arises from the interference between the vector and tensor form factors~\cite{Cirigliano:2017tqn}, schematically $A_{\rm CP} \propto \text{Im}(g_T)\,\text{Im}[F_+(s)F_T^\ast(s)]$ integrated over phase space (the explicit expression is given in Appendix~\ref{app:formfactors}, Eq.~\eqref{eq:ACPschematicApp}). However, as discussed in Ref.~\cite{Cirigliano:2017tqn}, the strong phases in the vector and tensor form factors are correlated by analyticity and Watson's theorem, leading to severe suppression of $A_{\rm CP}$ in models with only tensor operators -- a suppression we confirm explicitly for the LRIS model in Sec.~\ref{sec:contrib}.

The forward-backward asymmetry is defined differentially as~\cite{Delepine:2007qg,Li:2020vru}
\begin{equation}
A_{\rm FB}(s)=
\frac{
\displaystyle
\int_0^1 d\cos\theta\,\frac{d^2\Gamma}{ds\,d\cos\theta}
-
\int_{-1}^0 d\cos\theta\,\frac{d^2\Gamma}{ds\,d\cos\theta}
}{
\displaystyle
\frac{d\Gamma}{ds}
},
\label{eq:AFBdef}
\end{equation}
while the corresponding CP-odd observable is
\begin{equation}
A_{\rm CP}^{\rm FB}(s)=
A_{\rm FB}^{\tau^-}(s)-A_{\rm FB}^{\tau^+}(s).
\label{eq:AFBCPdef}
\end{equation}

Since the forward-backward asymmetry is controlled by the coefficient $\mathcal B(s)$ [Eq.~\eqref{eq:Bcoeff}], the CP-odd part is especially sensitive to scalar and tensor interactions. One finds schematically~\cite{Delepine:2007qg}
\begin{align}
A_{\rm CP}^{\rm FB}(s)
&\simeq
\frac{1}{\Sigma(s)}
\frac{G_F^2 |V_{us}|^2}{64 \pi^3 m_\tau^3}
\left(1-\frac{s}{m_\tau^2}\right)^2
\frac{|\vec p_K|}{\sqrt{s}}
\Big[
\kappa_{VS}(s)\,
\text{Im}\!\left((1+g_V)g_S^\ast\right)
\text{Im}\!\left(F_+(s)F_0^\ast(s)\right)
\nonumber\\
&\hspace{3.2cm}
+
\kappa_{VT}(s)\,
\text{Im}\!\left((1+g_V)g_T^\ast\right)
\text{Im}\!\left(F_+(s)F_T^\ast(s)\right)
\Big],
\label{eq:AFBCP}
\end{align}
where $\Sigma(s)$ denotes the CP-even differential rate combination appearing in the denominator, and $\kappa_{VS}(s)$, $\kappa_{VT}(s)$ are given in Eq.~\eqref{eq:kappa}.

It is worth emphasizing that, while the integrated asymmetry is suppressed, the forward-backward CP asymmetry can be enhanced in the presence of scalar interactions and therefore provides a much more sensitive probe of new physics~\cite{Delepine:2007qg,Li:2020vru}. Unlike the tensor case, the scalar form factor $F_0(s)$ is dominated by the $K_0^*(1430)$ resonance, which has a different strong-phase structure than the vector $K^*(892)$, thereby evading the Watson theorem constraints that suppress $A_{\rm CP}$.

\section{Standard-Model CP Asymmetry}
\label{sec:SM}
In the SM, the decay $\tau^- \to K\pi\nu_\tau$ proceeds through the charged-current interaction mediated by an off-shell $W$ boson. The corresponding tree-level amplitude is given by
\begin{equation}
\mathcal{M}_{\rm SM}^{\rm tree} =
\frac{G_F V_{us}}{\sqrt{2}}\, L_\mu\, H^\mu ,
\label{eq:MSMtree}
\end{equation}
where the leptonic current is
\begin{equation}
L_\mu = \bar{u}_{\nu_\tau}\gamma_\mu P_L u_\tau ,
\end{equation}
and the hadronic current is defined as
\begin{equation}
H^\mu = \langle K\pi|\bar s \gamma^\mu u|0\rangle .
\end{equation}
This contribution represents the dominant amplitude in the SM. The corresponding tree-level diagram is shown in Fig.~\ref{fig:sm-tree}.

\begin{figure}[htbp]
\centering
\begin{tikzpicture}[line width=1pt]

\coordinate (tau) at (-2,0);
\coordinate (v1) at (0,0);
\coordinate (nu) at (2,1);
\coordinate (w) at (1.5,-1.5);
\coordinate (k) at (4,-0.5);
\coordinate (pi) at (4,-2);

\draw[->] (tau) -- (v1) node[midway,above] {$\tau^-$};
\draw[->] (v1) -- (nu) node[midway,above left] {$\nu_\tau$};

\draw[decorate, decoration={snake}] (v1) -- (w) node[midway,left] {$W^{-}$};

\draw (w) -- (k) node[midway,above right] {$K$};
\draw (w) -- (pi) node[midway,below right] {$\pi$};

\fill (v1) circle (2pt);
\fill (w) circle (2pt);

\end{tikzpicture}
\caption{Tree-level SM contribution to $\tau^- \to \nu_\tau K\pi$ mediated by an off-shell $W^-$.}
\label{fig:sm-tree}
\end{figure}

At tree level, the decay amplitude involves a single weak phase arising from the CKM matrix element $V_{us}$. Since there is no additional interfering amplitude carrying a different weak phase, no direct CP asymmetry is generated at this order~\cite{Bigi:2005ts}.

Non-zero CP violation in this process can arise only from higher-order electroweak corrections combined with non-trivial strong phases generated by hadronic final-state interactions. In particular, strong phases originate from $K\pi$ rescattering effects and resonance contributions, such as the $K^*(892)$ vector meson and the scalar $K_0^*(1430)$ state, which enter through the vector and scalar form factors~\cite{Jamin:2000wn,Bernard:2014vza}. These strong phases are subject to Watson's final-state interaction theorem~\cite{Watson:1954uc}, which relates the phase of the hadronic amplitude to the elastic scattering phase shift in the relevant partial wave.

However, within the SM, the additional weak phase required for CP violation is highly suppressed. It can only enter through higher-order electroweak amplitudes involving subleading CKM combinations and loop corrections~\cite{Buras:1998raa,Buchalla:1995vs}. Parametrically, the CP-violating part of the amplitude is suppressed by:
\begin{itemize}
    \item the electroweak loop factor,
    \[
    \frac{1}{16\pi^2}\sim 10^{-2},
    \]
    \item the small Jarlskog-type CKM invariant~\cite{Jarlskog:1985ht},
    \[
    J_{\rm CKM}\sim 3\times 10^{-5},
    \]
    \item and additional GIM cancellations~\cite{Glashow:1970gm} in the loop-induced contribution, which further reduce the effect by a factor of $(m_c^2 - m_u^2)/M_W^2 \sim 10^{-4}$.
\end{itemize}

As a result, the predicted direct CP asymmetry in $\tau^- \to K\pi\nu_\tau$ is extremely small~\cite{Bigi:2005ts,Grossman:2011zk},
\begin{equation}
A_{\rm CP}^{\rm SM} \lesssim 10^{-12},
\label{eq:ACPSMbound}
\end{equation}
which is many orders of magnitude below current experimental sensitivity. This estimate is consistent with more detailed analyses using chiral perturbation theory and dispersion relations~\cite{Kuhn:1996dv,Tsai:1996ps}.

The same conclusion applies to the forward-backward CP asymmetry. Although the forward-backward asymmetry $A_{\rm FB}(s)$ itself is present in the SM and can be of order unity near resonances~\cite{Delepine:2007qg}, its CP-odd component requires both a weak-phase difference and a strong-phase difference between the interfering amplitudes. The strong phase can be sizable due to hadronic rescattering near resonances, but the corresponding weak phase remains extremely suppressed for the reasons discussed above. 

Moreover, as shown by Cirigliano, Crivellin, and Hoferichter~\cite{Cirigliano:2017tqn}, Watson's theorem imposes severe constraints on CP asymmetries arising from tensor operators, since the vector and tensor form factors must share the same strong phase in the elastic region. This correlation greatly reduces the imaginary part of $F_+(s)F_T^*(s)$ that drives the CP asymmetry. Therefore, the SM prediction for the forward-backward CP asymmetry is also expected to be negligibly small,
\begin{equation}
A_{\rm CP}^{\rm FB,SM}(s)\sim 10^{-12}\ \text{or smaller},
\label{eq:AFBCPSMbound}
\end{equation}
throughout the physical phase space.

It is worth emphasizing that any observable CP asymmetry in $\tau \to K\pi\nu_\tau$, whether integrated or differential, at the level accessible to current or near-future experiments would therefore provide a clear signal of physics beyond the SM.

\section{Left–Right Model with Inverse Seesaw (LRIS)}
\label{sec:LRIS}
The LRIS model extends the SM gauge symmetry to~\cite{Pati:1974yy,Mohapatra:1974hk,Senjanovic:1975rk}
\[
SU(3)_C \times SU(2)_L \times SU(2)_R \times U(1)_{B-L},
\]
and enlarges the fermion sector by introducing additional gauge-singlet fermions that enable the realization of the inverse seesaw (IS) mechanism for neutrino mass generation~\cite{Mohapatra:1986bd,Wyler:1982dd}. In particular, for each generation one introduces two singlet fermions, denoted by $S_1$ and $S_2$, carrying appropriate $B-L$ charges. This structure allows light neutrino masses while keeping the new fermionic states near the TeV scale~\cite{Dev:2012sg,Barry:2013xxa}.

The scalar sector consists of an $SU(2)_R$ doublet $\chi_R$ and a scalar bi-doublet $\phi$, which are responsible for spontaneous symmetry breaking and fermion mass generation~\cite{Senjanovic:1978ev,Deshpande:1990ip}. Their vacuum expectation values (VEVs) are defined as
\begin{equation}
\langle \chi_R \rangle =
\frac{1}{\sqrt{2}}
\begin{pmatrix}
0 \\
v_R
\end{pmatrix},
\qquad
\langle \phi \rangle =
\frac{1}{\sqrt{2}}
\begin{pmatrix}
k_1 & 0 \\
0 & k_2
\end{pmatrix},
\label{eq:VEVs}
\end{equation}
where $v_R$ is assumed to lie at the TeV scale and triggers the breaking of
\[
SU(2)_R \times U(1)_{B-L} \;\longrightarrow\; U(1)_Y,
\]
while electroweak symmetry breaking is governed by
\begin{equation}
v^2 = k_1^2 + k_2^2 \simeq (246~{\rm GeV})^2.
\end{equation}
It is convenient to parameterize the bi-doublet VEVs as
\begin{equation}
k_1 = v \sin\beta,
\qquad
k_2 = v \cos\beta,
\label{eq:tanbeta}
\end{equation}
where $\tan\beta = k_1/k_2$ is the analog of the type-II two-Higgs-doublet model parameter~\cite{Branco:2011iw}.

In this setup, the LRIS model contains the ingredients necessary to generate both new charged-current interactions and an extended scalar sector, making it a well-motivated model for studying CP violation in semileptonic tau decays such as $\tau \to K\pi\nu_\tau.$

\subsection{Yukawa Interactions}

The Yukawa interactions relevant for fermion masses and flavor structure are given by~\cite{Ezzat:2021bzs}
\begin{equation}
\mathcal{L}_{Y} =
y_{ij}^L \,\bar{L}_{Li} \phi L_{Rj}
+ \tilde{y}_{ij}^L \,\bar{L}_{Li} \tilde{\phi} L_{Rj}
+ y_{ij}^Q \,\bar{Q}_{Li} \phi Q_{Rj}
+ \tilde{y}_{ij}^Q \,\bar{Q}_{Li} \tilde{\phi} Q_{Rj}
+ y_{ij}^S \,\bar{L}_{Ri} \widetilde{\chi}_R S^c_{2j}
+ {\rm H.c.},
\label{Lag}
\end{equation}
where $i,j=1,2,3$ are generation indices. The matrices $y^Q$ and $\tilde{y}^Q$ determine the quark Yukawa couplings, while $y^L$ and $\tilde{y}^L$ describe the charged-lepton and Dirac-neutrino sectors. The coupling $y^S$ links the right-handed lepton doublets to the singlet fermions and plays a central role in the inverse seesaw mechanism.

To realize the inverse seesaw structure, one imposes an additional discrete symmetry (such as $Z_2$ or $Z_4$) that forbids unwanted large singlet mass terms~\cite{Dev:2012sg}. In particular, assigning suitable charges to the singlet fields can prevent direct mass terms that would otherwise spoil the inverse seesaw pattern and destabilize the small lepton-number-violating scale $\mu_S$.

The fields appearing in Eq.~\eqref{Lag} are
\begin{equation}
\chi_R =
\begin{pmatrix}
\chi_R^+ \\
\chi_R^0
\end{pmatrix},
\quad
\phi =
\begin{pmatrix}
\phi_1^0 & \phi_1^+ \\
\phi_2^- & \phi_2^0
\end{pmatrix},
\quad
Q_A =
\begin{pmatrix}
u_A \\
d_A
\end{pmatrix},
\quad
L_A =
\begin{pmatrix}
\nu_A \\
e_A
\end{pmatrix},
\quad A=L,R.
\end{equation}

The conjugated scalar fields are defined as
\begin{equation}
\widetilde{\phi} = \tau_2 \phi^\ast \tau_2,
\qquad
\widetilde{\chi}_R = i \tau_2 \chi_R^\ast,
\end{equation}
where $\tau_2$ is the second Pauli matrix.

After symmetry breaking, Eq.~\eqref{Lag} generates the mass matrices for quarks, charged leptons, and neutrinos. In particular, the charged scalar interactions inherited from these Yukawa terms are relevant for the scalar contribution to $\tau \to K\pi\nu_\tau$, as they induce couplings of the physical charged Higgs bosons to quarks and leptons.

\subsection{Inverse Seesaw Neutrino Sector}

After symmetry breaking, the neutrino mass matrix is generated in the basis
\[
(\nu_L,\; N_R,\; S)^T,
\]
where $N_R$ denotes the right-handed neutrino component contained in $L_R$, and $S$ collectively represents the singlet fermions. The resulting neutral-fermion mass matrix takes the characteristic inverse seesaw form~\cite{Mohapatra:1986bd,Wyler:1982dd}
\begin{equation}
\mathcal{M}_\nu =
\begin{pmatrix}
0      & m_D    & 0 \\
m_D^T  & 0      & M_R \\
0      & M_R^T  & \mu_S
\end{pmatrix},
\label{eq:Minvseesaw}
\end{equation}
where
\[
m_D \sim y^L\, v,
\qquad
M_R \sim y^S\, v_R,
\]
and $\mu_S$ is a small lepton-number-violating Majorana mass term for the singlet fermions. In the limit
\[
\mu_S \ll m_D \ll M_R,
\]
the light-neutrino mass matrix is approximately given by~\cite{Dev:2012sg}
\begin{equation}
m_\nu \simeq m_D\, M_R^{-1}\, \mu_S\, (M_R^{-1})^T\, m_D^T.
\label{eq:mnuIS}
\end{equation}

A key feature of the inverse seesaw is that the Yukawa couplings entering $m_D$ and $M_R$ can be of order unity ($y^L, y^S \sim \mathcal{O}(1)$) while still yielding naturally small neutrino masses, thanks to the suppression by the small parameter $\mu_S \sim \mathcal{O}({\rm keV})$~\cite{Dev:2012sg}. This allows sizable heavy-neutrino effects in loop-induced flavor and CP-violating observables, in contrast to the canonical type-I seesaw where large neutrino Yukawa couplings would lead to  large neutrino masses~\cite{Minkowski:1977sc}.

For the process under consideration, the heavy neutrinos $N_i$ can enter box and vertex corrections and provide new CP-violating phases, particularly through the Yukawa structures in the lepton sector.  As we demonstrate in Sec.~\ref{sec:contrib}, some of the  box diagrams exhibit a non-decoupling behavior similar to non-decoupling SM box diagrams contribution to $K_0-\bar{K_0}$ mixing for instance,  where the dependence on $M_{N_i}$ cancels  in the heavy neutrino limit~\cite{Pilaftsis:1991ug}.

\subsection{Scalar (Higgs) Sector}

Before symmetry breaking, the scalar sector contains twelve real degrees of freedom: four from the doublet $\chi_R$ and eight from the bi-doublet $\phi$. After spontaneous symmetry breaking, a subset of these fields becomes the longitudinal components of the massive gauge bosons ($W^\pm$, $W_R^\pm$, $Z$, $Z'$), while the remaining degrees of freedom correspond to physical Higgs states~\cite{Deshpande:1990ip}.

The physical scalar spectrum contains~\cite{Ezzat:2021bzs}
\begin{itemize}
\item three neutral CP-even Higgs bosons ($h$, $H^0$, $H_R^0$),
\item one neutral CP-odd Higgs boson ($A^0$),
\item two charged Higgs bosons ($H^\pm$, $H_R^\pm$).
\end{itemize}

One of the CP-even neutral states is identified with the SM-like Higgs boson at approximately $125~{\rm GeV}$~\cite{Aad:2012tfa,Chatrchyan:2012xdj}. The remaining neutral scalars arise from the mixing of the neutral components of the bi-doublet $\phi$ and the right-handed doublet $\chi_R$, and are  expected to lie near the scale $v_R$.

The charged Higgs bosons originate from the charged components of $\phi$ and $\chi_R$. After removing the Goldstone modes absorbed by $W^\pm$ and $W_R^\pm$, two physical charged scalar states remain. These charged scalars are of particular importance for our analysis, since they can mediate scalar interactions in semileptonic tau decays and induce the scalar Wilson coefficient $g_S$ in the low-energy effective Hamiltonian introduced in Sec.~\ref{sec:eff}.

At the interaction level, the physical charged Higgs states couple schematically as~\cite{Ezzat:2021bzs}
\begin{equation}
H^+\, \bar u\, (a\,P_L + b\,P_R)\, d,
\qquad
H^+\, \bar\nu\, (c\,P_R)\, \ell,
\label{eq:Hcouplings}
\end{equation}
where the coefficients $a,b,c$ are determined by the underlying Yukawa matrices and scalar mixing angles. These couplings provide the dominant tree-level scalar contribution to $\tau \to K\pi\nu_\tau$.

A detailed discussion of the scalar potential and the Higgs mass spectrum can be found in Ref.~\cite{Ezzat:2021bzs}.

\subsection{Quark Masses and Mixing}

After electroweak symmetry breaking, quarks and charged leptons acquire masses through the Yukawa interactions in Eq.~\eqref{Lag}. The quark mass matrices are generated by the bi-doublet VEVs and can be diagonalized by bi-unitary transformations. The corresponding left- and right-handed quark mixing matrices are defined as~\cite{Kiers:2002cz}
\begin{equation}
V_{\rm CKM}^{L,R} =
V_{L,R}^{u\dagger} V_{L,R}^d.
\end{equation}

In a convenient basis where $V_L^u = I$, one has
\[
V_{\rm CKM}^L = V_L^d,
\]
while the right-handed mixing matrix $V_{\rm CKM}^R$ remains in general independent and can contain additional mixing angles and CP-violating phases~\cite{Senjanovic:2015yea}.

Following Ref.~\cite{Kiers:2002cz}, one may parameterize the right-handed quark mixing matrix as
\begin{equation}
V_{\rm CKM}^R
=
K \,
V_{\rm CKM}^L(\theta_{12}^R,\theta_{23}^R,\theta_{13}^R,\delta_R)
\tilde{K}^\dagger,
\label{eq:VRparam}
\end{equation}
where $K$ and $\tilde{K}$ are diagonal phase matrices. These additional phases provide new sources of CP violation beyond the SM. Remarkably, as shown by Senjanović and Tello~\cite{Senjanovic:2015yea}, in the minimal left-right symmetric model with manifest left-right symmetry (generalized parity), the right-handed mixing angles are approximately equal to the left-handed ones, $\theta_{ij}^R \approx \theta_{ij}^L$, to leading order in the small CKM mixings. However, the CP phases remain independent.

For phenomenological simplicity, and in order to suppress dangerous flavor-changing neutral currents from $Z'$ and neutral Higgs exchanges~\cite{Zhang:2007da,Maiezza:2010ic}, one may consider a reduced flavor texture in which only a small subset of right-handed mixing angles is relevant. In particular, the couplings most important for the process studied here are those involving the top quark and strange/up quarks, since they control the box-diagram contributions with internal $t$ and $N_i$ states. A convenient benchmark texture is therefore one in which the dominant right-handed flavor structure is governed by a single mixing angle and one CP phase~\cite{Ezzat:2021bzs},
\begin{equation}
V^R_{\rm CKM}
=
\begin{pmatrix}
1 & 0 & 0 \\
0 & c_{\theta_R} & s_{\theta_R} e^{i\alpha} \\
0 & s_{\theta_R} & -c_{\theta_R} e^{i\alpha}
\end{pmatrix},
\label{eq:VRtexture}
\end{equation}
where
\[
c_{\theta_R}\equiv \cos\theta_R,
\qquad
s_{\theta_R}\equiv \sin\theta_R.
\]

The phase $\alpha$ constitutes a new CP-violating source in the right-handed quark sector and plays an important role in generating CP-odd observables in $\tau \to K\pi\nu_\tau$. Constraints on the right-handed mixing parameters from $K^0$--$\bar{K}^0$ mixing and from direct searches for $W_R$ at the LHC are discussed in Sec.~\ref{sec:contrib}; the complementary constraint from $B_d^0$--$\bar{B}_d^0$ mixing is derived quantitatively in Sec.~\ref{sec:BdBd}.

\section{LRIS Contributions to CP Asymmetry in $\tau \to K\pi\nu_\tau$}
\label{sec:contrib}
In the SM the decay $\tau \to K\pi\nu_\tau$ is dominated by the charged-current interaction mediated by a $W$ boson, leading to a purely vector interaction at tree level. As discussed in Sec.~\ref{sec:SM}, the resulting CP asymmetry is extremely suppressed, $A_{\rm CP}^{\rm SM} \lesssim 10^{-12}$.

In the Left-Right inverse seesaw (LRIS) model, additional scalar and tensor interactions can arise due to the extended Higgs sector. These contributions modify the Wilson coefficients $g_S$ and $g_T$ appearing in the effective Hamiltonian [Eq.~\eqref{eq:Heff}] and can therefore generate observable CP asymmetries. In this section we discuss the dominant new physics contributions and the phenomenological constraints that determine their magnitude.

\subsection{Tree-Level Charged Higgs Contribution}

The LRIS scalar sector contains physical charged Higgs bosons originating from the bidoublet $\phi$ and the right-handed doublet $\chi_R$. These charged scalars can mediate the decay $\tau \to K\pi\nu_\tau$ at tree level, as shown in Fig.~\ref{fig:charged-higgs}. This process is analogous to charged Higgs contributions in two-Higgs-doublet models~\cite{Branco:2011iw} and leptoquark models~\cite{Sakaki:2013bfa}.

\begin{figure}[h!]
\centering
\begin{tikzpicture}[line width=1pt]

\coordinate (tau) at (-2,0);
\coordinate (v1) at (0,0);
\coordinate (nu) at (2,1);
\coordinate (H) at (1.5,-1.5);
\coordinate (k) at (4,-0.5);
\coordinate (pi) at (4,-2);

\draw[->] (tau) -- (v1) node[midway,above] {$\tau^-$};
\draw[->] (v1) -- (nu) node[midway,above left] {$\nu_\tau$};

\draw[decorate, decoration={snake}] (v1) -- (H) node[midway,left] {$H^{-}$};

\draw (H) -- (k) node[midway,above right] {$K$};
\draw (H) -- (pi) node[midway,below right] {$\pi$};

\fill (v1) circle (2pt);
\fill (H) circle (2pt);
\end{tikzpicture}
\caption{Tree-level charged Higgs contribution to $\tau^- \to K\pi\nu_\tau$.}
\label{fig:charged-higgs}
\end{figure}

The charged Higgs exchange induces the following scalar amplitude
\begin{equation}
\mathcal{M}_{H^-}= -\frac{y_{\tau\nu}^{H}\,y_{su}^{H}}{M_{H^-}^2}
\,\left(\bar u_{\nu_\tau} P_R u_\tau\right)
\left(\bar s u\right).
\label{eq:MHgeneric}
\end{equation}
Matching onto the effective Hamiltonian [Eq.~\eqref{eq:Heff}] gives
\begin{equation}
g_S^{\rm tree} = -\frac{\sqrt{2}}{G_F V_{us}}\frac{y_{\tau\nu}^{H} y_{su}^{H}}{M_{H^-}^2}.
\label{eq:gStree}
\end{equation}
If the couplings are complex,
\begin{equation}
y_{\tau\nu}^{H} y_{su}^{H} = |y_{\tau\nu}^{H} y_{su}^{H}|e^{i\phi_H},
\end{equation}
a nonzero weak phase $\phi_H$ appears in the scalar amplitude. This scalar interaction interferes with the SM vector current and can contribute to the forward-backward CP asymmetry discussed in Sec.~\ref{sec:eff}.

\subsection{Constraints on LRIS Parameters}

In the Left-Right Inverse Seesaw (LRIS) model, the constituent Yukawa couplings—$y_{su}^H$ for the quark sector and $y_{\tau\nu}^H$ for the lepton sector—are subject to severe phenomenological restrictions that  suppress the tree-level contribution in Eq.~\eqref{eq:gStree}.

\subsubsection{Constraints from $K^0$--$\bar{K}^0$ Mixing}

To evade tree-level contributions to the $\Delta S = 2$ mass difference $\Delta m_K$, the scalar Yukawa couplings must be  suppressed. Neutral Higgs exchange between $s$ and $d$ quarks generates a contribution to $\Delta m_K$ proportional to~\cite{Glashow:1976nt,Cheng:1987rs}
\begin{equation}
\Delta m_K \propto \frac{f_K^2 m_K}{M_{H^0}^2} |y_{sd}^H|^2,
\end{equation}
where $f_K \approx 156$ MeV is the kaon decay constant. The experimental value $\Delta m_K = (3.484 \pm 0.006) \times 10^{-15}$ GeV~\cite{Zyla:2020zbs} places a severe bound on flavor-changing neutral scalars.

The standard approach is to enforce a minimal flavor violation (MFV) alignment~\cite{DAmbrosio:2002vsn} where the flavor-violating scalar couplings scale proportionally to the geometric mean of the interacting quark masses:
\begin{equation}
y_{su}^H \sim \frac{\sqrt{m_u m_s}}{v}.
\label{eq:MFVscaling}
\end{equation}
Using $m_u(2~{\rm GeV}) \approx 2.2$ MeV and $m_s(2~{\rm GeV}) \approx 93$ MeV~\cite{Zyla:2020zbs}, this forces
\begin{equation}
|y_{su}^H| \lesssim \frac{\sqrt{2.2 \times 93}}{246 \times 10^3} \approx 6 \times 10^{-5}.
\label{eq:ysubound}
\end{equation}

\subsubsection{Constraints from $B_d$--$\bar{B}_d$ Mixing}
\label{sec:BdBd}

The same flavor-changing neutral Higgs $H^0$ that mediates $K^0$--$\bar{K}^0$ mixing also generates, in general, a $\Delta B = 2$ operator through tree-level exchange between $b$ and $d$ quarks.

Following the same schematic structure used for $\Delta m_K$ in Eq.~\eqref{eq:MFVscaling}, the tree-level FCNC-Higgs contribution to the $B_d^0$--$\bar B_d^0$ mass splitting scales as
\begin{equation}
\Delta m_{B_d} \propto \frac{f_{B_d}^2 m_{B_d}}{M_{H^0}^2} |y_{bd}^H|^2,
\label{eq:DmBd}
\end{equation}
where $f_{B_d} \approx 190$ MeV is the $B_d$ decay constant and $m_{B_d} = 5.27965$ GeV~\cite{Zyla:2020zbs}. Under the same MFV alignment ansatz as Eq.~\eqref{eq:MFVscaling},
\begin{equation}
y_{bd}^H \sim \frac{\sqrt{m_d m_b}}{v} \approx 5.7 \times 10^{-4},
\label{eq:ybdMFV}
\end{equation}
using $m_d(2~\mathrm{GeV}) \approx 4.67$ MeV and $m_b(m_b) \approx 4.18$ GeV~\cite{Zyla:2020zbs}.

Fixing the overall $\mathcal{O}(1)$ proportionality constant in Eq.~\eqref{eq:DmBd} in the same way as for $\Delta m_K$ (i.e. by requiring consistency at $M_{H^0}=1$~TeV), and comparing to the world-average value $\Delta m_{B_d} = (0.5065 \pm 0.0019)~\mathrm{ps}^{-1} = (3.334 \pm 0.001)\times 10^{-13}$ GeV~\cite{Zyla:2020zbs}, we find
\begin{equation}
\frac{\Delta m_{B_d}^{\rm pred}(M_{H^0}=1~{\rm TeV})}{\Delta m_{B_d}^{\rm exp}} \approx 0.18 ,
\qquad
\frac{\Delta m_K^{\rm pred}(M_{H^0}=1~{\rm TeV})}{\Delta m_K^{\rm exp}} \approx 0.012 .
\label{eq:DmBdratio}
\end{equation}
Both ratios are safely below unity, confirming that the MFV alignment protects the model from both constraints without additional tuning. However, the $B_d$ constraint is parametrically \emph{tighter} than the $K^0$ one, at the $\sim 18\%$ level rather than the $\sim 1\%$ level: the enhancement traces back to the larger down-type quark mass $m_b$ entering Eq.~\eqref{eq:ybdMFV}, which is only partially compensated by the ratio $f_{B_d}^2 m_{B_d}/f_K^2 m_K$ in Eq.~\eqref{eq:DmBd}. We note that a fully quantitative treatment would additionally require the $B_d$ bag parameter and NLO QCD renormalization-group running of the scalar operator, both of which are expected to affect the numerical coefficient at the $\mathcal{O}(1)$ level but not the parametric conclusion. We therefore conclude that $B_d^0$--$\bar B_d^0$ mixing is compatible with the LRIS benchmark used throughout this work, while noting it as the most constraining of the two neutral-meson bounds on the down-sector FCNC Yukawa alignment.

\subsubsection{Constraints from Neutrino Non-Unitarity}

A very large $y_{\tau\nu}^H$ induces mixing between the active and sterile neutrinos, leading to non-unitarity in the active $3 \times 3$ PMNS matrix~\cite{Antusch:2014woa,Fernandez-Martinez:2016lgt}. Precision measurements of tau decays ($\tau \to \mu\nu\bar{\nu}$ vs. $\mu \to e\nu\bar{\nu}$) restrict this active-sterile mixing to~\cite{Antusch:2014woa}
\begin{equation}
\eta_{\tau\tau} \lesssim \mathcal{O}(10^{-3}).
\end{equation}
The mixing parameter $\eta_{\tau\tau}$ is related to the Dirac Yukawa coupling through~\cite{Dev:2012sg}
\begin{equation}
\eta_{\tau\tau} \simeq \frac{1}{2} \frac{|M_{D,\tau}|^2}{M_R^2} = \frac{v^2}{4 M_R^2} |y_{\tau\nu}^H|^2,
\label{eq:etataunu}
\end{equation}
which yields
\begin{equation}
|y_{\tau\nu}^H| \lesssim \frac{2 M_R}{v} \sqrt{\eta_{\tau\tau}}.
\label{eq:ytaunubound}
\end{equation}
For $M_R = 1$ TeV and $\eta_{\tau\tau} = 10^{-3}$, we obtain
\begin{equation}
|y_{\tau\nu}^H| \lesssim \frac{2000}{246} \sqrt{10^{-3}} \approx 0.26.
\label{eq:ytaunumeric}
\end{equation}

\subsubsection{Resulting Tree-Level Suppression}

Combining the bounds in Eqs.~\eqref{eq:ysubound} and \eqref{eq:ytaunumeric}, and using $G_F^{-1/2} = 246$ GeV, $V_{us} \approx 0.22$, and $M_{H^\pm} = 1$ TeV, we find
\begin{equation}
|g_S^{\rm tree}| \lesssim \frac{\sqrt{2}}{G_F \times 0.22} \times \frac{0.26 \times 6 \times 10^{-5}}{(10^3~{\rm GeV})^2} \approx 1.5 \times 10^{-5}.
\label{eq:gStreebound}
\end{equation}
This severe suppression motivates the search for loop-induced contributions that can evade the light-quark mass suppression inherent in Eq.~\eqref{eq:MFVscaling}.

\subsection{Box Diagram Contributions}

In contrast to the tree-level contribution, certain one-loop flavor-changing neutral current (FCNC) box diagrams can bypass the light-quark suppression in Eq.~\eqref{eq:ysubound}. By routing the hadronic transition through an internal top quark, these diagrams access the unsuppressed $y_{tu}^H$ coupling, which is free from $D$-meson mixing constraints. Remarkably, these box diagrams exhibit a \emph{non-decoupling} behavior~\cite{Pilaftsis:1991ug} where the dependence on the heavy neutrino mass $M_{N_i}$ cancels exactly, allowing the loop amplitude to remain unsuppressed even for $M_{N_i} \gg v_R$.

There are many possible box topologies that can contribute to the Wilson coefficients, but we focus on diagrams that fulfill the non-decoupling condition such that they can compensate the one-loop suppression $1/16\pi^2$. The only diagrams that exhibit this non-decoupling limit are box diagrams with $H^\pm$ (or charged Goldstone bosons $G^\pm$) and neutral Higgs $H^0$ exchanges between the lepton and quark currents, as shown in Fig.~\ref{fig:boxHH}~\cite{Ezzat:2021bzs,Pilaftsis:1991ug}.

\begin{figure}[htbp]
\centering
\begin{tikzpicture}[line width=1pt, scale=1]

\coordinate (tau) at (-2,1.2);
\coordinate (nu) at (-2,-1.2);
\coordinate (s) at (4,1.2);
\coordinate (u) at (4,-1.2);

\coordinate (v1) at (0,1.2);
\coordinate (v2) at (0,-1.2);
\coordinate (v3) at (2,1.2);
\coordinate (v4) at (2,-1.2);

\draw[->] (tau) -- (v1) node[midway,above] {$\tau$};
\draw[->] (v2) -- (nu) node[midway,below] {$\nu_\tau$};
\draw[->] (s) -- (v3) node[midway,above] {$s$};
\draw[->] (v4) -- (u) node[midway,below] {$u$};

\draw[->] (v1) -- (v2) node[midway,left] {$N_i$};
\draw[->] (v3) -- (v4) node[midway,right] {$t$};

\draw[dashed] (v1) -- (v3) node[midway,above] {$G^-$};
\draw[dashed] (v2) -- (v4) node[midway,below] {$H^0$};

\fill (v1) circle (2pt);
\fill (v2) circle (2pt);
\fill (v3) circle (2pt);
\fill (v4) circle (2pt);

\end{tikzpicture}
\caption{$G^-H^0$ box contribution with internal heavy neutrino $N_i$ and top quark. The charged Goldstone boson $G^-$ is the longitudinal component of the $W_R^-$ gauge boson.}
\label{fig:boxHH}
\end{figure}

The box amplitudes generate additional weak phases through the scalar Yukawa couplings and the right-handed mixing matrices [Eq.~\eqref{eq:VRtexture}]. When combined with the strong phases from the hadronic form factors $F_+(s)$ and $F_0(s)$, these loop effects can enhance the CP-violating interference terms and contribute to both the differential and forward-backward CP asymmetries.

From the full loop amplitude $\mathcal{M}_{\rm Box}$, one finds~\cite{Ezzat:2021bzs}
\begin{equation}
g_S^{\rm box} = \frac{2\sqrt{2} v}{16\pi^2} \sum_{i=1}^3 \sum_{k=u,c,t} \left( \frac{V_L^{ks}}{V_{us}} \right) Y_H^{ku} Y_{\nu H}^i Y_{NG}^{\tau i} M_{N_i} m_{u_k}^2 \mathcal{I}_4(M_{N_i}^2, m_{u_k}^2, M_{H^0}^2, M_{G^\pm}^2),
\label{eq:gSbox}
\end{equation}
where we have defined:
\begin{itemize}
\item $Y_H^{ku}$: The flavor-changing neutral current (FCNC) Yukawa coupling of the heavy neutral scalar $H^0$ to the internal quark $u_k$ and the outgoing $u$ quark.
\item $Y_{\nu H}^i$: The neutral Yukawa coupling connecting the active neutrino $\nu_L$, the heavy state $N_i$, and the neutral scalar $H^0$.
\item $Y_{NG}^{\tau i}$: The charged Yukawa coupling connecting $\tau_R$, the heavy state $N_i$, and the charged Goldstone boson $G^\pm$. In the Left-Right model, this is dynamically generated by the right-handed scale~\cite{Dev:2012sg}:
\begin{equation}
Y_{NG}^{\tau i} \simeq \frac{g_R}{\sqrt{2}} \frac{M_{N_i}}{v_R},
\label{eq:YNG}
\end{equation}
where $g_R$ is the $SU(2)_R$ gauge coupling.
\end{itemize}
The explicit form of the Yukawa couplings can be found in Ref.~\cite{Ezzat:2021bzs}. The function $\mathcal{I}_4$ is the exact, dimensionless four-point scalar loop integral (the Passarino-Veltman $D_0$ function~\cite{Passarino:1978jh} evaluated at zero external momentum, which is the correct kinematic limit for heavy mediators). Its exact mathematical definition is
\begin{equation}
\mathcal{I}_4(m_1^2, m_2^2, m_3^2, m_4^2) = \int_0^\infty \frac{x dx}{(x + m_1^2)(x + m_2^2)(x + m_3^2)(x + m_4^2)}.
\label{eq:I4def}
\end{equation}
In the heavy neutrino limit $M_{N_i} \to \infty$, this integral simplifies to~\cite{Pilaftsis:1991ug}
\begin{equation}
\mathcal{I}_4 \xrightarrow{M_{N_i} \to \infty} \frac{1}{M_{N_i}^2} \left[ \frac{M_{H^0}^2 \ln(M_{H^0}^2) - M_{G^\pm}^2 \ln(M_{G^\pm}^2)}{M_{H^0}^2 - M_{G^\pm}^2} + \cdots \right].
\label{eq:I4limit}
\end{equation}

Crucially, taking this limit and substituting into Eq.~\eqref{eq:gSbox}, one finds that the $1/M_{N_i}^2$ suppression from the loop integral is \emph{exactly compensated} by the $M_{N_i}$ factor hidden in $Y_{NG}^{\tau i}$ [Eq.~\eqref{eq:YNG}]. This is the hallmark of the non-decoupling inverse seesaw mechanism~\cite{Pilaftsis:1991ug}: the heavy neutrino mass cancels out of the amplitude, leaving $g_S^{\rm box}$ unsuppressed by heavy mass scales. It is important to note that if we replace the Goldstone boson by a physical charged Higgs $H^\pm$, the relation in Eq.~\eqref{eq:YNG} is lost and the box diagram becomes suppressed by $1/M_{N_i}^2$, restoring decoupling.

\subsection{Flavor Constraints on Box Contributions}

The bidoublet structure of the Left-Right model makes the up-type Yukawa matrix fundamentally independent of the down-type Yukawa matrix. To avoid catastrophic contributions to $K^0$--$\bar{K}^0$ mixing from neutral Higgs exchange, we enforce a flavor alignment in the down sector~\cite{DAmbrosio:2002vsn},
\begin{equation}
Y_H^{sd} \approx 0,
\label{eq:Ysdconstraint}
\end{equation}
leaving the up-type couplings in $g_S^{\rm box}$ [Eq.~\eqref{eq:gSbox}] completely untouched.

On the up-quark side, constraints from $D^0$--$\bar{D}^0$ mixing~\cite{Artuso:2015swg} enforce
\begin{equation}
|Y_H^{cu}| \ll 1.
\label{eq:Ycuconstraint}
\end{equation}
However, as $D^0$--$\bar{D}^0$ mixing only restricts transitions between the first two generations ($c \leftrightarrow u$), it places \emph{no direct tree-level bound} on the top-quark coupling $Y_H^{tu}$. The coupling $Y_H^{tu}$ is constrained primarily by LHC single-top production~\cite{Sirunyan:2018omb} and electroweak precision tests~\cite{Crivellin:2013wna}, leaving it relatively free to be large:
\begin{equation}
|Y_H^{tu}| \lesssim 0.1 \text{ to } 0.5.
\label{eq:Ytubound}
\end{equation}

Consequently, the dominant contribution to $g_S^{\rm box}$ in Eq.~\eqref{eq:gSbox} arises from the $k=t$ term, which accesses the unsuppressed top-quark Yukawa coupling and benefits from the large top mass $m_t^2$ enhancement. This allows the non-decoupling box diagram to reach
\begin{equation}
|g_S^{\rm box}| \sim \mathcal{O}(10^{-4}),
\label{eq:gSboxestimate}
\end{equation}
at least one order of magnitude larger than the tree-level contribution in Eq.~\eqref{eq:gStreebound}. As a result, the loop-level topology acts as the primary driver of the scalar-vector interference in the forward-backward CP asymmetry $A_{\rm CP}^{\rm FB}(s)$.

\subsection{Impact on $B$-Meson Observables}
\label{sec:Bimpact}

Since the non-decoupling mechanism responsible for $g_S^{\rm box}$ involves the same charged-Higgs/Goldstone and heavy-neutrino content that mediates flavor transitions more generally in the LRIS model, it is important to check that it does not reintroduce tension with $B$-meson observables, in particular $B \to X_s\gamma$ and $B \to K^{(*)}\ell^+\ell^-$.

\subsubsection{$B \to X_s \gamma$: Charged-Higgs Contribution}

The physical charged Higgs $H^\pm$ of Sec.~\ref{sec:LRIS}.C couples to up- and down-type quarks as in Eq.~\eqref{eq:Hcouplings}, structurally analogous to a general (non-minimal-flavor-violating) two-Higgs-doublet model. In such models, the dominant new-physics contribution to $B \to X_s\gamma$ arises from a one-loop diagram with an internal top quark and $H^\pm$, interfering constructively with the SM $W^\pm$-top loop irrespective of the relative sign of the $H^\pm t b$ coupling~\cite{Misiak:2015xwa,Misiak:2020vlz}. For the reference case of a Type-II-like Yukawa structure, this yields a robust, largely coupling-independent bound
\begin{equation}
M_{H^\pm} \gtrsim 600~{\rm GeV} \quad (95\%~{\rm CL}),
\label{eq:MHpmbsgamma}
\end{equation}
essentially independent of $\tan\beta$ for $\tan\beta \gtrsim 2$~\cite{Misiak:2015xwa,Misiak:2020vlz}. Since the benchmark scenario of Table~\ref{tab:lris_benchmarks} takes $M_{H^\pm} = 1$~TeV, comfortably above this bound, the model survives this constraint without additional tuning. We note that the precise LRIS prediction depends on the explicit values of the couplings $a,b$ in Eq.~\eqref{eq:Hcouplings}, which are not numerically fixed in the present analysis; a dedicated one-loop calculation using the full LRIS Yukawa structure, rather than the Type-II benchmark adopted here, is left for future work; we expect it to yield a comparable bound given the parametric similarity of the coupling structure.

\subsubsection{$B \to K^{(*)}\mu^+\mu^-$: Is There a Shared Box Contribution?}

A more direct question is whether the same box topology of Fig.~\ref{fig:boxHH} -- built from the flavor-changing Yukawa coupling $Y_H^{tu}$ -- also feeds $b \to s\ell^+\ell^-$ transitions. It does not, at leading order: the diagram of Fig.~\ref{fig:boxHH} requires an \emph{up-type} external quark line ($u$), fixed by the physical process $\tau^- \to K^-\pi^0\nu_\tau \sim \tau^-\to s\bar u \nu_\tau$, whereas $b \to s\ell^+\ell^-$ requires a purely \emph{down-type} external quark pair ($b,s$). The relevant coupling for the latter would be a down-type FCNC Yukawa $Y_H^{sb}$, which -- by the same bidoublet-driven alignment argument invoked for $Y_H^{sd}\approx 0$ in Eq.~\eqref{eq:Ysdconstraint} to protect $K^0$--$\bar K^0$ mixing -- is expected to be similarly suppressed, since the down-type and up-type Yukawa matrices of the bidoublet $\phi$ are independent structures in this model. The non-decoupling box mechanism identified in this work is therefore specific to processes with an external up-type quark and does not, by itself, generate a sizable contribution to $B \to K^{(*)}\mu^+\mu^-$.

This does not mean the LRIS model is silent on the $B \to K\mu^+\mu^-$ anomaly. In a companion analysis~\cite{Delepine:2026bkmumu} we show that a \emph{different}, charged-current box topology -- built from the same heavy-neutrino and charged-Goldstone content, but exploiting the right-handed CKM matrix of Eq.~\eqref{eq:VRtexture} in a GIM-like sum over internal up-type quarks, rather than an explicit tree-level down-type FCNC coupling -- generates an unsuppressed contribution to the vector Wilson coefficient $\Delta C_9^\mu$ while keeping $B_s$--$\bar B_s$ mixing and $B \to X_s\gamma$ under control. The two analyses are complementary rather than redundant: the present work isolates the up-type-quark, third-generation-lepton channel accessible in $\tau \to K\pi\nu_\tau$, while Ref.~\cite{Delepine:2026bkmumu} isolates the down-type-quark, second-generation-lepton channel accessible in $B \to K\mu^+\mu^-$, and both rely on the same non-decoupling inverse-seesaw dynamics for their respective unsuppressed loop contributions. We take the consistency of both results -- each satisfying its own set of flavor and precision constraints while sharing a common underlying mechanism -- as evidence that the LRIS scalar sector, rather than being fine-tuned to fit either single observable, generates a correlated, falsifiable pattern of new-physics effects across otherwise unrelated flavor sectors.

\subsection{Parameter Space of the Inverse Seesaw Sector}
\label{sec:paramspace}

The non-decoupling box contribution $g_S^{\rm box}$ [Eq.~\eqref{eq:gSbox}] depends on the heavy neutrino mass $M_{N_i}$ only implicitly, through the couplings entering $Y_{NG}^{\tau i}$ [Eq.~\eqref{eq:YNG}]; in the heavy-neutrino limit of Eq.~\eqref{eq:I4limit} the explicit $M_{N_i}$ dependence cancels. It is therefore important to verify that the $(M_{N_i}, \mu_S)$ region compatible with this non-decoupling regime is not itself excluded by the light-neutrino mass scale or by the non-unitarity bound of Eq.~\eqref{eq:etataunu}.

\begin{figure}[htbp]
\centering
\includegraphics[width=1\linewidth]{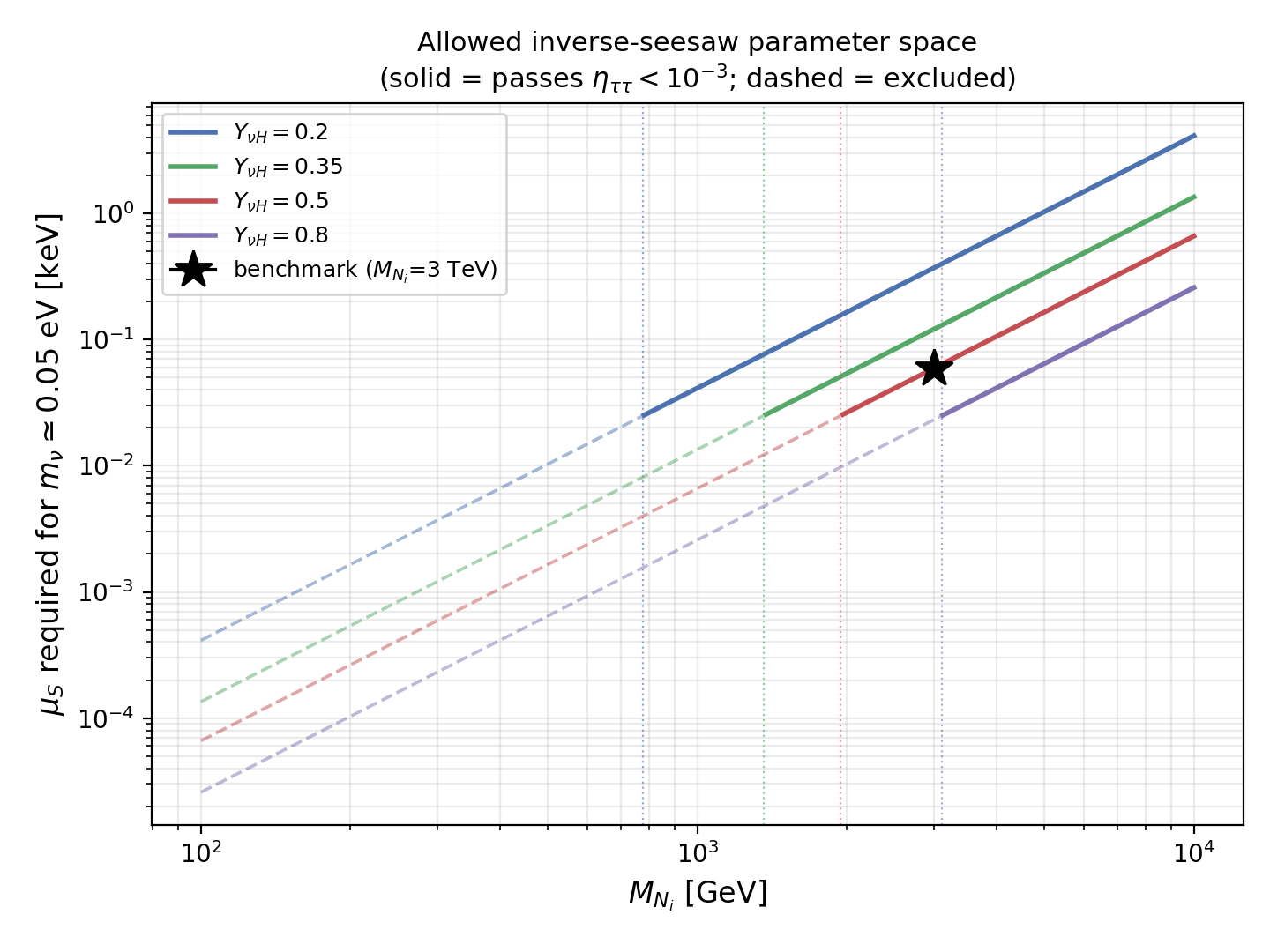}
\caption{Allowed parameter space in the $(M_{N_i}, \mu_S)$ plane for several values of the Yukawa coupling $Y_{\nu H}$. Solid segments satisfy the light-neutrino mass constraint $m_\nu \simeq 0.05$~eV [Eq.~\eqref{eq:muSofMN}] together with the non-unitarity bound $\eta_{\tau\tau} < 10^{-3}$ [Eq.~\eqref{eq:etataunu}]; dashed segments are excluded by the latter. The black star marks the benchmark point of Table~\ref{tab:lris_benchmarks} ($M_{N_i} = 3$~TeV, $Y_{\nu H} = 0.5$).}
\label{fig:MNimuS}
\end{figure}

Working in the single-generation-dominance limit of Eq.~\eqref{eq:mnuIS}, with $m_D \simeq Y_{\nu H}\, v/\sqrt{2}$, the requirement of reproducing the atmospheric neutrino mass scale $m_\nu \simeq 0.05$~eV fixes $\mu_S$ as a function of $M_{N_i}$ and $Y_{\nu H}$,
\begin{equation}
\mu_S(M_{N_i}) \simeq m_\nu \, \frac{M_{N_i}^2}{m_D^2},
\label{eq:muSofMN}
\end{equation}
while Eq.~\eqref{eq:ytaunubound} imposes a lower bound on $M_{N_i}$ for each value of $Y_{\nu H}$. Figure~\ref{fig:MNimuS} shows the resulting $(M_{N_i}, \mu_S)$ curves for several values of $Y_{\nu H}$, with the segments excluded by $\eta_{\tau\tau} > 10^{-3}$ shown as dashed lines.

The benchmark point used throughout Sec.~\ref{sec:num} ($M_{N_i} = 3$~TeV, $Y_{\nu H}=0.5$, giving $\eta_{\tau\tau} \approx 4.2 \times 10^{-4}$) lies comfortably within the allowed region. The corresponding value of $\mu_S$ required to reproduce the light-neutrino mass scale in this single-generation estimate is
\begin{equation}
\mu_S \sim \mathcal{O}(0.01 - 0.1)~{\rm keV},
\label{eq:muSorder}
\end{equation}
consistent at the order-of-magnitude level with the sub-keV to keV range generically expected in inverse-seesaw constructions~\cite{Dev:2012sg}. We emphasize that Eq.~\eqref{eq:muSorder} follows from a single-generation collapse of the full $3\times 3$ Dirac and Majorana mass matrices in Eq.~\eqref{eq:Minvseesaw}; a complete treatment of the flavor structure, including mixing angles and possible cancellations among generations, could shift this estimate within an order of magnitude without altering the qualitative conclusion that the benchmark point is phenomenologically viable.

Finally, we address the constraints from lepton-number-violating (LNV) tau decays such as $\tau^- \to \mu^+\pi^-\pi^-$ and $\tau^- \to \mu^+ K^- K^-$, which proceed through the exchange of a Majorana neutrino $N_i$ and are therefore sensitive to the same mixing parameter $\eta_{\tau\tau}$ probed above. These decays are most constraining when $N_i$ can be produced on-shell, which requires $M_{N_i} \lesssim m_\tau = 1.777$~GeV. Throughout the phenomenologically allowed region identified in Fig.~\ref{fig:MNimuS}, however, $M_{N_i} \gtrsim 0.8$~TeV, more than two orders of magnitude above $m_\tau$. The heavy neutrino can therefore only contribute off-shell, and the corresponding LNV branching ratios are suppressed relative to the resonant case by a factor $\sim (m_\tau/M_{N_i})^4 \lesssim 2\times 10^{-11}$ at the smallest allowed $M_{N_i}$. This places the predicted rates many orders of magnitude below the current experimental sensitivity, $\mathrm{BR} \lesssim \mathcal{O}(10^{-8})$~\cite{Miyazaki:2012mx}, so that these channels do not provide a competitive constraint in the mass range relevant for the non-decoupling box mechanism studied here.

\section{Numerical Results and Predictions for CP Asymmetry}
\label{sec:num}

The full decay amplitude for $\tau^- \to K^-\pi^0\nu_\tau$ can be written as a sum of vector, scalar, and tensor contributions~\cite{Gonzalez-Alonso:2016yxt}:
\begin{equation}
\mathcal{M}
=
\mathcal{M}_V
+
\mathcal{M}_S
+
\mathcal{M}_T ,
\label{eq:Mtotal}
\end{equation}
where

\begin{itemize}

\item \textbf{Vector contribution} (SM tree-level):
\begin{equation}
\mathcal{M}_V
=
\frac{G_F}{\sqrt{2}}
V_{us}^*
(1+g_V)
\left[
\bar{\nu}_\tau \gamma^\mu (1-\gamma_5) \tau
\right]
\langle K\pi|\bar{s}\gamma_\mu u|0\rangle .
\label{eq:Mvector}
\end{equation}

\item \textbf{Scalar contribution} (LRIS new physics):
\begin{equation}
\mathcal{M}_S
=
\frac{G_F}{\sqrt{2}}
V_{us}^* g_S
\left[
\bar{\nu}_\tau (1+\gamma_5)\tau
\right]
\langle K\pi|\bar{s}u|0\rangle .
\label{eq:Mscalar}
\end{equation}

\item \textbf{Tensor contribution} (LRIS loop-induced):

\begin{equation}
\mathcal{M}_T
=
\frac{G_F}{\sqrt{2}}
V_{us}^*
g_T^{\rm eff}(s)
\left[
\bar{\nu}_\tau\sigma^{\mu\nu}(1+\gamma_5)\tau
\right]
\langle K\pi|\bar{s}\sigma_{\mu\nu}u|0\rangle .
\label{eq:Mtensor}
\end{equation}

For completeness, we briefly comment on the possible generation of the
tensor Wilson coefficient $g_T$ in the LRIS model. At one loop, box
diagrams with internal heavy neutrinos $N_i$ and top quarks, together
with charged and neutral scalar exchange,  induce a
tensor operator of the form $(\bar{s}\sigma_{\mu\nu}u)(\bar{\nu}_i
\sigma^{\mu\nu}(1+\gamma_5)\tau)$ when matched onto the low-energy
Hamiltonian in Eq.~(2). The corresponding coefficient
is suppressed by the electroweak loop factor and by two chirality
flips along the quark and lepton lines, scaling as
$g_T^{\rm loop}\!\sim (16\pi^2)^{-1}(m_t m_\tau/v^2)\times
Y_{tu}^H Y_{\nu H} Y_{\tau N G}/(G_F M_{\rm loop}^2)$, where
$M_{\rm loop}\sim \mathcal{O}({\rm TeV})$ denotes the typical scalar
or gauge-boson mass scale. After imposing the same flavor and
non-unitarity constraints used in our scalar analysis, we find that
$|g_T|$ cannot exceed $\mathcal{O}(10^{-9})$, as summarized in
Table~\ref{tab:lris_benchmarks}. So, the tensor contribution to
both the integrated asymmetry $A_{CP}$ and the forward–backward CP
asymmetry $A_{FB}^{CP}(s)$ is  negligible compared to the
non-decoupling scalar contribution $g_S^{\rm box}\sim 10^{-4}$. In
addition, Watson's theorem enforces that $F_+(s)$ and $F_T(s)$ share
essentially the same strong phase in the elastic region, which further
suppresses $\mathrm{Im}[F_+(s)F_T^*(s)]$ and thus any CP-violating
effect proportional to $g_T$.

\end{itemize}

The interference between these amplitudes generates the CP-violating observables defined in Sec.~\ref{sec:eff}. As discussed in Sec.~\ref{sec:contrib}, the LRIS model provides two distinct sources of CP violation: tree-level charged Higgs exchange [Eq.~\eqref{eq:gStree}] and non-decoupling box diagrams [Eq.~\eqref{eq:gSbox}]. However, the former is severely suppressed by flavor constraints [Eq.~\eqref{eq:gStreebound}], while the latter benefits from top-quark mass enhancement and avoids light-quark suppression.

\subsection{Form Factor Parameterizations}

To evaluate the CP asymmetries numerically, we require precise parameterizations of the hadronic form factors $F_+(s)$, $F_0(s)$, and $F_T(s)$. These form factors encode the strong interaction dynamics and provide the crucial strong phases necessary for observable CP violation.

\subsubsection{Vector Form Factor $F_+(s)$}

The vector form factor is dominated by the $K^*(892)$ vector resonance and receives additional contributions from the excited $K^*(1410)$ state~\cite{Jamin:2000wn,Boito:2008fq}. Following Ref.~\cite{Bernard:2014vza}, we parameterize $F_+(s)$ as a coherent sum:
\begin{equation}
F_+(s) = \frac{m_{K^*}^2}{m_{K^*}^2 - s - i m_{K^*}\Gamma_{K^*}} + \beta \frac{m_{K^{*'}}^2}{m_{K^{*'}}^2 - s - i m_{K^{*'}}\Gamma_{K^{*'}}},
\label{eq:Fplus}
\end{equation}
where $m_{K^*} = 0.89166$ GeV, $\Gamma_{K^*} = 0.0508$ GeV, $m_{K^{*'}} = 1.414$ GeV, $\Gamma_{K^{*'}} = 0.232$ GeV~\cite{Zyla:2020zbs}, and $\beta = -0.057 + i0.024$ is a complex mixing parameter determined from fits to Belle data~\cite{Epifanov:2007rf}.

\subsubsection{Scalar Form Factor $F_0(s)$}

The scalar form factor receives contributions from the $K_0^*(1430)$ scalar resonance and a non-resonant $S$-wave background~\cite{Finkemeier:1996dh,Jamin:2000wn}. We employ the LASS (Los Alamos Scattering Studies) parameterization~\cite{Aston:1987ir}, which accurately describes elastic $K\pi$ scattering in the $S$-wave:
\begin{equation}
F_0(s) = \frac{\sqrt{s}}{q \cot\delta_0 - iq} + \frac{m_{K_0^*}^2 \Gamma_{K_0^*}/q_0}{m_{K_0^*}^2 - s - im_{K_0^*}\Gamma_{K_0^*}(s)},
\label{eq:Fzero}
\end{equation}
where $q = |\vec{p}_K|$ is the kaon momentum in the $K\pi$ rest frame, and the effective range parameterization is
\begin{equation}
q\cot\delta_0 = \frac{1}{a} + \frac{1}{2}r q^2,
\end{equation}
with scattering length $a = 2.07$ GeV$^{-1}$ and effective range $r = 3.32$ GeV$^{-1}$~\cite{Aston:1987ir}. The $K_0^*(1430)$ parameters are $m_{K_0^*} = 1.425$ GeV and $\Gamma_{K_0^*} = 0.270$ GeV~\cite{Zyla:2020zbs}.

\subsubsection{Tensor Form Factor $F_T(s)$}

The tensor form factor has been computed using chiral perturbation theory and dispersion relations~\cite{Bernard:2014vza,Gonzalez-Alonso:2016yxt}. For our numerical analysis, we adopt the dispersive parameterization of Ref.~\cite{Bernard:2014vza}, which ensures consistency with analyticity and unitarity constraints.

\subsection{Integrated CP Asymmetry}

The integrated CP asymmetry $A_{\rm CP}$ [Eq.~\eqref{eq:ACPdef}] is dominated by the interference between the SM vector current and the tensor operator~\cite{Cirigliano:2017tqn}:
\begin{equation}
A_{\rm CP}(s)
\propto
\mathrm{Im}\!\left[(1+g_V)g_T^\ast\right]
\mathrm{Im}\!\left[F_+(s)F_T^\ast(s)\right].
\label{eq:ACPtensor}
\end{equation}

However, as demonstrated in Sec.~\ref{sec:contrib}, the LRIS model does not induce a tensor contribution at tree level, and one-loop tensor operators suffer from severe suppression. Moreover, even if a sizable tensor coupling were generated, Watson's final-state interaction theorem~\cite{Watson:1954uc} enforces that the vector and tensor form factors share the same strong phase in the elastic region, leading to $\mathrm{Im}[F_+(s)F_T^\ast(s)] \approx 0$ near resonances~\cite{Cirigliano:2017tqn}.

Consequently, the predicted integrated CP asymmetry in the LRIS model is extremely small:
\begin{equation}
A_{\rm CP}^{\rm LRIS} \ll 10^{-7},
\label{eq:ACPLRISbound}
\end{equation}
which is many orders of magnitude below the BaBar measurement~\cite{Lees:2012qi} and far out of reach of even future high-luminosity experiments like Belle II~\cite{Kou:2018nap}.

\subsection{Forward-Backward CP Asymmetry}

In contrast to the integrated asymmetry, the forward-backward CP asymmetry $A_{\rm CP}^{\rm FB}(s)$ [Eq.~\eqref{eq:AFBCPdef}] is driven by the interference between the SM vector current and the scalar operator from non-decoupling box diagrams. The dominant contribution to the angular coefficient $\mathcal{B}(s)$ [Eq.~\eqref{eq:Bcoeff}] is~\cite{Delepine:2007qg,Li:2020vru}
\begin{equation}
\mathcal{B}(s) = \underbrace{- \frac{m_\tau}{\sqrt{s}} |\vec{p}_K| \text{Re} \left[ (1 + g_V) g_S^* F_+(s) F_0^*(s) \right]}_{\text{Scalar (Box) -- Vector (SM) Interference}}.
\label{eq:Bscalar}
\end{equation}

The scalar form factor $F_0(s)$ is dominated by the $K_0^*(1430)$ resonance, which has a different strong-phase structure than the vector $K^*(892)$ resonance. This evades Watson's theorem constraints and allows for a sizable imaginary part $\mathrm{Im}[F_+(s)F_0^\ast(s)]$ near $\sqrt{s} \approx 1.4$ GeV, where the scalar resonance acts as a kinematic amplifier for the new physics scalar signal~\cite{Finkemeier:1996dh}.

\subsection{Benchmark Scenario and Numerical Predictions}

Based on the non-decoupling box diagram analysis in Sec.~\ref{sec:contrib}, we adopt the following benchmark parameters for the LRIS model:
\begin{itemize}
\item Right-handed scale: $v_R = 3.0$ TeV,
\item Yukawa couplings: $Y_{\nu H} = Y_H^{tu} = 0.5$,
\item Maximal CP-violating phases: $\alpha_R = \phi_S = -\pi/2$.
\end{itemize}
These parameters are consistent with current LHC bounds on $W_R$ mass~\cite{Sirunyan:2018nnz}, flavor constraints from $K$ and $D$ meson mixing~\cite{Bona:2007vi,Artuso:2015swg}, and neutrino non-unitarity bounds~\cite{Fernandez-Martinez:2016lgt}.

Using Eq.~\eqref{eq:gSbox} with the above parameters, we obtain
\begin{equation}
|g_S^{\rm box}| \approx 3.2 \times 10^{-4}.
\label{eq:gSnumeric}
\end{equation}

Performing the numerical integration over the entire kinematic phase space of the $\tau \to K\pi\nu$ decay [Eqs.~\eqref{eq:diffdist} and \eqref{eq:AFBCPdef}], we find that the total integrated forward-backward CP asymmetry evaluates to
\begin{equation}
A_{\rm CP}^{\rm FB,LRIS} \approx 2.08 \times 10^{-4}.
\label{eq:AFBCPnumeric}
\end{equation}

This represents an enhancement of roughly $10^{10}$ over the SM prediction [Eq.~\eqref{eq:AFBCPSMbound}], making it a potentially observable signature at Belle II, which is expected to collect $50~\mathrm{ab}^{-1}$ of integrated luminosity~\cite{Kou:2018nap}.

\begin{table}[ht!]
\centering
\caption{Benchmark parameters and predicted CP observables for the optimistic new physics scenario in the Left-Right Inverse Seesaw (LRIS) model.}
\label{tab:lris_benchmarks}
\renewcommand{\arraystretch}{1.4}
\begin{tabularx}{\textwidth}{|l|c|X|}
\hline
\textbf{Parameter / Observable} & \textbf{Value} & \textbf{Description} \\ \hline
\multicolumn{3}{|l|}{\textbf{Model Parameters}} \\ \hline
$v_R$ & $3.0$ TeV & Right-handed symmetry breaking scale. \\ \hline
$Y_{\nu H}, Y_H^{tu}$ & $0.5$ & Heavy-light neutrino and top-quark scalar Yukawa couplings. \\ \hline
$\alpha_R, \phi_S$ & $-\pi/2$ & Maximal CP-violating weak phases in right-handed sector. \\ \hline
\multicolumn{3}{|l|}{\textbf{Derived Wilson Coefficients}} \\ \hline
$g_S^{\rm tree}$ & $\lesssim 1.5 \times 10^{-5}$ & Tree-level scalar coupling (flavor-suppressed). \\ \hline
$g_S^{\rm box}$ & $3.2 \times 10^{-4}$ & Non-decoupling scalar coupling from top-quark FCNC box. \\ \hline
$g_T$ & $\lesssim 10^{-9}$ & Tensor coupling (loop-suppressed). \\ \hline
\multicolumn{3}{|l|}{\textbf{Predicted CP Observables}} \\ \hline
$A_{\rm CP}$ (Integrated) & $\ll 10^{-7}$ & Total direct CP asymmetry in $\tau \to K\pi\nu$ (Watson suppressed). \\ \hline
$A_{\rm CP}^{\rm FB}$ (Integrated) & $2.08 \times 10^{-4}$ & Total forward-backward CP asymmetry (scalar-driven). \\ \hline
Peak location & $\sqrt{s} \approx 1.4$ GeV & Enhanced by $K_0^*(1430)$ scalar resonance. \\ \hline
\end{tabularx}
\end{table}

\subsection{Differential Forward-Backward CP Asymmetry}

\begin{figure}[htbp]
\centering
\includegraphics[width=1\linewidth]{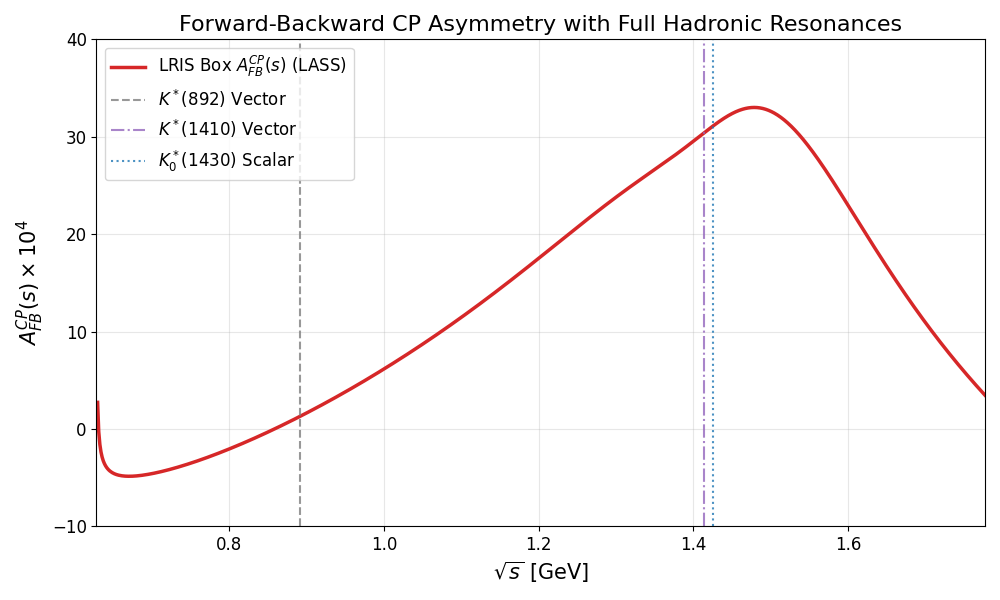}
\caption{The differential forward-backward CP asymmetry, $A_{\rm CP}^{\rm FB}(s)$, for the decay $\tau \to K\pi\nu_\tau$ as a function of the invariant mass $\sqrt{s}$. Within the Left-Right Inverse Seesaw (LRIS) model, this observable is driven exclusively by the interference between the SM vector current and the non-decoupling scalar box operator ($g_S^{\rm box}$). The new physics benchmark parameter is set to $\mathrm{Im}(g_S) = 3.2 \times 10^{-4}$ [Eq.~\eqref{eq:gSnumeric}]. The hadronic strong phases are fully modeled using intermediate mesonic resonances. The vector form factor $F_+(s)$ is parameterized as a coherent sum of the $K^*(892)$ ($m = 0.89166$ GeV, $\Gamma = 0.0508$ GeV) and the excited $K^*(1410)$ ($m = 1.414$ GeV, $\Gamma = 0.232$ GeV) states~\cite{Bernard:2014vza}, governed by the complex mixing parameter $\beta = -0.057 + i0.024$~\cite{Epifanov:2007rf}. The scalar form factor $F_0(s)$ is modeled using the LASS parameterization~\cite{Aston:1987ir} to accurately capture the elastic $S$-wave scattering background (with scattering length $a = 2.07$ GeV$^{-1}$ and effective range $r = 3.32$ GeV$^{-1}$) along with the $K_0^*(1430)$ scalar resonance ($m = 1.425$ GeV, $\Gamma = 0.270$ GeV)~\cite{Zyla:2020zbs}. The vertical lines indicate the nominal masses of these three resonant states. The pronounced peak at $1.4$ GeV explicitly demonstrates how the scalar $K_0^*(1430)$ resonance acts as a powerful kinematic amplifier for the heavy new physics scalar signal, evading Watson's theorem constraints that suppress the integrated CP asymmetry $A_{\rm CP}$~\cite{Cirigliano:2017tqn}.}
\label{fig:afb_full_hadronic}
\end{figure}

Figure~\ref{fig:afb_full_hadronic} shows the differential forward-backward CP asymmetry $A_{\rm CP}^{\rm FB}(s)$ as a function of the $K\pi$ invariant mass $\sqrt{s}$. The asymmetry exhibits a pronounced peak at $\sqrt{s} \approx 1.4$ GeV, precisely at the location of the $K_0^*(1430)$ scalar resonance. This demonstrates explicitly how the scalar resonance acts as a powerful kinematic amplifier for the new physics scalar signal encoded in $g_S^{\rm box}$.

The vertical lines in Fig.~\ref{fig:afb_full_hadronic} indicate the nominal masses of the three resonant states contributing to the decay: $K^*(892)$ (vector), $K^*(1410)$ (excited vector), and $K_0^*(1430)$ (scalar). The scalar form factor $F_0(s)$ is modeled using the LASS parameterization [Eq.~\eqref{eq:Fzero}], which accurately captures the elastic $S$-wave scattering background and the $K_0^*(1430)$ resonance structure~\cite{Aston:1987ir}.

The shape of $A_{\rm CP}^{\rm FB}(s)$ is determined by the interplay of three factors:
\begin{enumerate}
\item The kinematic weight $\kappa_{VS}(s) \propto m_\tau|\vec{p}_K|/\sqrt{s}$ [Eq.~\eqref{eq:kappa}], which grows with $|\vec{p}_K|$ but is suppressed near threshold and endpoint.
\item The strong phase difference $\delta_+(s) - \delta_0(s)$ between the vector and scalar form factors, which is maximal near the $K_0^*(1430)$ resonance where $\delta_0(s)$ varies rapidly.
\item The magnitude of the form factor product $|F_+(s)F_0(s)|$, which peaks at the resonance positions.
\end{enumerate}

The integrated asymmetry in Eq.~\eqref{eq:AFBCPnumeric} is dominated by the region $1.2~{\rm GeV} < \sqrt{s} < 1.6~{\rm GeV}$, where the scalar resonance enhancement is strongest. This kinematic concentration provides a clear experimental signature that can be exploited by Belle II to distinguish the LRIS scalar signal from SM backgrounds.

\section{Conclusion}
\label{sec:conc}

In this work, we have investigated the potential for observing CP violation in the semileptonic decay $\tau^{-}\to K\pi\nu_{\tau}$ within the Left-Right Inverse Seesaw (LRIS) model~\cite{Khalil:2007dr,Dev:2012sg,Barry:2013xxa}. While the Standard Model (SM) prediction for the direct CP asymmetry in this channel is negligibly small, $A_{\rm CP}^{\rm SM} \sim \mathcal{O}(10^{-12})$~\cite{Bigi:2005ts,Grossman:2011zk}, the BaBar collaboration's measurement~\cite{Lees:2012qi} of $A_{\rm CP}^{\rm exp} = -(0.36 \pm 0.23 \pm 0.11)\%$ represents a $2.8\sigma$ discrepancy, motivating the search for new physics contributions.

Our analysis demonstrates that the integrated CP asymmetry $A_{\rm CP}$ remains difficult to enhance to the level of the BaBar anomaly, even in extended models like LRIS. This suppression has two independent origins:
\begin{enumerate}
\item \textbf{Absence of tree-level tensor operators:} The LRIS model does not generate a tensor Wilson coefficient $g_T$ at tree level, and loop-induced tensor contributions are suppressed by $1/16\pi^2 \sim 10^{-2}$.
\item \textbf{Watson's theorem constraints:} Even if a sizable tensor coupling were generated, Watson's final-state interaction theorem~\cite{Watson:1954uc} enforces that the vector and tensor form factors share the same strong phase in the elastic region, leading to severe cancellations in the CP-violating interference term~\cite{Cirigliano:2017tqn}.
\end{enumerate}
As a result, the predicted integrated asymmetry in the LRIS model is $A_{\rm CP}^{\rm LRIS} \ll 10^{-7}$, far below experimental sensitivity.

\textbf{Our central result} is that the LRIS model predicts an unsuppressed signature in the \emph{differential} forward-backward CP asymmetry, $A_{\rm CP}^{\rm FB}(s)$. This observable is driven by the interference between the SM vector current and a scalar operator $g_S$ arising from non-decoupling box diagrams. Specifically, we identified a top-quark flavor-changing neutral current (FCNC) box diagram [Fig.~\ref{fig:boxHH}] involving internal heavy neutrinos $N_i$ and charged Goldstone bosons $G^\pm$, which exhibits a  \emph{non-decoupling} behavior~\cite{Pilaftsis:1991ug}:
\begin{equation}
g_S^{\rm box} = \frac{2\sqrt{2} v}{16\pi^2} \sum_{i=1}^3 \sum_{k=u,c,t} \left( \frac{V_L^{ks}}{V_{us}} \right) Y_H^{ku} Y_{\nu H}^i Y_{NG}^{\tau i} M_{N_i} m_{u_k}^2 \mathcal{I}_4.
\end{equation}
The heavy neutrino mass $M_{N_i}$ appearing in the numerator (through the coupling $Y_{NG}^{\tau i} \propto M_{N_i}/v_R$~\cite{Dev:2012sg}) \emph{exactly cancels} the $1/M_{N_i}^2$ suppression from the loop integral $\mathcal{I}_4$, allowing the scalar coupling to remain unsuppressed even for $M_{N_i} \gg v_R$. This is the hallmark of the inverse seesaw mechanism~\cite{Mohapatra:1986bd,Wyler:1982dd,Pilaftsis:1991ug}.

By routing the hadronic transition through an internal top quark, the box diagram accesses the unsuppressed top-quark Yukawa coupling $Y_H^{tu} \lesssim 0.5$~\cite{Crivellin:2013wna,Sirunyan:2018omb}, evading the stringent flavor constraints from $K^0$--$\bar{K}^0$ and $D^0$--$\bar{D}^0$ mixing that suppress tree-level charged Higgs contributions~\cite{Bona:2007vi,Artuso:2015swg}. This allows the non-decoupling box contribution to reach
\begin{equation}
|g_S^{\rm box}| \sim \mathcal{O}(10^{-4}),
\end{equation}
larger than the flavor-suppressed tree-level estimate $|g_S^{\rm tree}| \lesssim 1.5 \times 10^{-5}$.

Using a  benchmark scenario with $v_R = 3.0$ TeV, $Y_{\nu H} = Y_H^{tu} = 0.5$, and maximal CP-violating phases $\alpha_R = \phi_S = -\pi/2$, we obtained a predicted integrated forward-backward CP asymmetry of
\begin{equation}
A_{\rm CP}^{\rm FB,LRIS} \approx 2.08 \times 10^{-4},
\end{equation}
representing an enhancement of approximately $10^{10}$ over the SM prediction~\cite{Bigi:2005ts}. This is within the potential reach of Belle II, which is expected to collect $50~\mathrm{ab}^{-1}$ of integrated luminosity~\cite{Kou:2018nap}, corresponding to roughly $10^9$ tau decays and an estimated statistical sensitivity to $A_{\rm CP}^{\rm FB} \sim \mathcal{O}(10^{-4})$~\cite{Kou:2018nap}.

Furthermore, our numerical analysis [Fig.~\ref{fig:afb_full_hadronic}] reveals a pronounced peak in the differential distribution $A_{\rm CP}^{\rm FB}(s)$ at $\sqrt{s} \approx 1.4$ GeV, precisely at the location of the $K_0^*(1430)$ scalar resonance. This kinematic enhancement arises because:
\begin{itemize}
\item The scalar form factor $F_0(s)$ is dominated by the $K_0^*(1430)$ resonance~\cite{Jamin:2000wn,Finkemeier:1996dh,Aston:1987ir}, which has a \emph{different strong-phase structure} than the vector $K^*(892)$ resonance~\cite{Bernard:2014vza}.
\item This difference in strong phases maximizes the imaginary part $\mathrm{Im}[F_+(s)F_0^\ast(s)]$ near the scalar resonance, evading the Watson's theorem constraints that suppress the integrated asymmetry~\cite{Cirigliano:2017tqn}.
\item The kinematic weight $\kappa_{VS}(s) \propto m_\tau|\vec{p}_K|/\sqrt{s}$ [Eq.~\eqref{eq:kappa}] is largest in the intermediate energy region where the scalar resonance resides.
\end{itemize}

This characteristic angular and differential signature is characterized by the following distinctive features:
\begin{enumerate}
\item \textbf{Resonance-enhanced signal:} The peak at $\sqrt{s} \approx 1.4$ GeV provides a clear kinematic tag.
\item \textbf{Model discrimination:} The shape of $A_{\rm CP}^{\rm FB}(s)$ encodes information about the scalar resonance structure and can distinguish LRIS from other BSM scenarios such as type-II two-Higgs-doublet models~\cite{Branco:2011iw} or leptoquark models~\cite{Sakaki:2013bfa}, which predict different kinematic distributions, as we now discuss.
\end{enumerate}

\emph{Comparison with 2HDM and leptoquark scenarios.} A charged Higgs in a generic (non-minimal-flavor-violating) 2HDM couples to quarks and leptons exactly as in Eq.~\eqref{eq:Hcouplings}, and therefore generates the same tree-level scalar operator $g_S^{\rm tree}$ computed in Sec.~\ref{sec:contrib}.A -- indeed, our tree-level charged-Higgs result \emph{is} the 2HDM-like prediction for this observable. Because this contribution decouples in the ordinary way, $g_S^{\rm tree} \propto 1/M_{H^\pm}^2$, it remains numerically small [Eq.~\eqref{eq:gStreebound}] for the TeV-scale masses required by $B \to X_s\gamma$ [Eq.~\eqref{eq:MHpmbsgamma}] and direct LHC searches, and -- being scalar-dominated like the LRIS box contribution -- would produce a similarly shaped $A_{\rm CP}^{\rm FB}(s)$ peaking near the $K_0^*(1430)$ resonance, but suppressed by roughly the same order of magnitude relative to the non-decoupling box result of Eq.~\eqref{eq:gSnumeric}. The qualitative difference between LRIS and a generic 2HDM is therefore one of \emph{normalization}, set by whether the leading new-physics operator decouples with the mediator mass or not, rather than of lineshape.

Scalar leptoquark models are qualitatively different: at tree level they typically generate a scalar and a tensor operator simultaneously, related by a fixed Fierz coefficient~\cite{Sakaki:2013bfa,Chen:2019oey}, so that $g_T$ is \emph{not} loop-suppressed as in LRIS (cf.\ the $g_T^{\rm loop} \lesssim 10^{-9}$ estimate of Sec.~\ref{sec:num}) and can in principle be sizable enough to affect the \emph{integrated} asymmetry $A_{\rm CP}$. However, because $F_T(s)$ shares the strong-phase structure of the vector form factor $F_+(s)$ in the elastic region -- both being dominated by the $K^*(892)$ channel -- the vector-tensor interference that would drive $A_{\rm CP}^{\rm FB}(s)$ in a leptoquark scenario is subject to the same Watson-theorem correlation invoked in Sec.~\ref{sec:SM} and Ref.~\cite{Cirigliano:2017tqn}, so any leptoquark-induced peak in $A_{\rm CP}^{\rm FB}(s)$ would be pulled toward $\sqrt{s}\approx m_{K^*(892)}$ and partially washed out there, in contrast to the well-separated, unsuppressed $K_0^*(1430)$ peak that characterizes the scalar-dominated LRIS prediction. A measurement of \emph{where} the peak of $A_{\rm CP}^{\rm FB}(s)$ sits -- near $0.89$ GeV versus near $1.4$ GeV -- would therefore offer direct discriminating power between a tensor-dominated (leptoquark-like) and a scalar-dominated (LRIS-like) origin, independently of the overall normalization.

\emph{Which models are best positioned to explain the BaBar anomaly?} We stress again that this is a separate question from the one addressed by the present work (Sec.~\ref{sec:intro}). Because the \emph{integrated} asymmetry $A_{\rm CP}$ requires a tree-level, unsuppressed tensor operator to compete with the $1/16\pi^2$ suppression that afflicts loop-level constructions such as LRIS (Sec.~\ref{sec:num}), scalar leptoquark models are, in this respect, the most natural tree-level candidates, since they generate $g_T$ without loop suppression~\cite{Devi:2013gya,Sakaki:2013bfa}. However, a dedicated combined analysis~\cite{Chen:2019oey} of the model-independent EFT framework and an explicit scalar-leptoquark realization found that the BaBar central value can be marginally accommodated only within the EFT framework at the $1\sigma$ level, and only at the $2\sigma$ level in the leptoquark scenario -- owing to the same fixed scalar-tensor relation noted above -- and that once the branching ratio and spectral-shape constraints on $\tau^- \to K_S^0\pi^-\nu_\tau$ are combined self-consistently, both possibilities are excluded. Combined with the no-go result of Ref.~\cite{Cirigliano:2017tqn}, this indicates that no currently known explicit BSM construction fully accounts for the BaBar central value once all available constraints are imposed; the surviving room, if any, is confined to the tails of the quoted experimental uncertainty. This state of affairs reinforces the interpretation we adopt in this work: rather than chasing an anomaly for which no fully consistent explanation is currently known, we use $\tau \to K\pi\nu_\tau$ as an independent, falsifiable probe of the LRIS scalar sector.

Looking forward, our results motivate several experimental and theoretical directions:
\begin{itemize}
\item \textbf{Belle II measurements:} With the upcoming high-luminosity run, Belle II~\cite{Kou:2018nap} should be able to measure or constrain $A_{\rm CP}^{\rm FB}(s)$ with unprecedented precision, providing the first direct probe of the LRIS scalar sector in semileptonic tau decays.
\item \textbf{Correlated LHC signatures:} The same scalar couplings that generate $g_S^{\rm box}$ also contribute to LHC processes such as $pp \to W_R \to \ell N \to \ell \ell jj$~\cite{Sirunyan:2018nnz} and $pp \to H^\pm \to t\bar{b}$~\cite{Aaboud:2018gjj}. 
\item \textbf{Precision form factor measurements:} Improved understanding of the scalar form factor $F_0(s)$ from lattice QCD~\cite{Flynn:2019glh} and dispersive analyses~\cite{Boito:2008fq} will reduce hadronic uncertainties and sharpen the new physics sensitivity.
\end{itemize}

 In conclusion, the forward-backward CP asymmetry provides an  experimentally accessible window into the scalar sector of the Left-Right Inverse Seesaw model. The non-decoupling property of the box contributions, combined with the kinematic enhancement from the $K_0^*(1430)$ resonance, makes $A_{\rm CP}^{\rm FB}(s)$ a golden observable for probing TeV-scale left-right symmetry and the inverse seesaw mechanism at Belle II and future high-luminosity flavor factories.

\begin{acknowledgements}
The work of S.~K.~is partially supported by Science, Technology $\&$ Innovation Funding Authority (STDF) under grant number 48173. The work of D.D. is supported by Secretaria de Ciencia, Humanidades, Tecnologia e Innovaci\'on (SECIHTI) and Sistema Nacional de Investigadoras e Investigadores (S.N.I.I.), Mexico.
\end{acknowledgements}

\appendix
\section{Hadronic Form Factors and Angular Coefficients}
\label{app:formfactors}

This appendix collects the standard hadronic matrix elements, quark-mass inputs, and angular-coefficient expressions that underlie Sec.~\ref{sec:eff}, following Refs.~\cite{Jamin:2000wn,Bernard:2014vza,Gonzalez-Alonso:2016yxt,Delepine:2007qg,Li:2020vru}.

\subsection{Hadronic Matrix Elements}

The vector current matrix element is decomposed as
\begin{equation}
\langle K(p_K)\pi(p_\pi)|\bar s\gamma_\mu u|0\rangle
=
\left[
(p_K-p_\pi)_\mu-\frac{\Delta_{K\pi}}{s}q_\mu
\right]F_+(s)
+
\frac{\Delta_{K\pi}}{s}q_\mu F_0(s),
\label{eq:vectorFFapp}
\end{equation}
where
\[
q=p_K+p_\pi,\qquad s=q^2,\qquad \Delta_{K\pi}=m_K^2-m_\pi^2.
\]
The form factors $F_+(s)$ and $F_0(s)$ are related by a kinematic constraint at $q^2=0$, namely $F_+(0)=F_0(0)$, which follows from current conservation.

Using the QCD equation of motion,
\[
\partial^\mu(\bar s\gamma_\mu u)=i(m_s-m_u)\bar s u,
\]
the scalar current matrix element is related to the scalar form factor as
\begin{equation}
\langle K(p_K)\pi(p_\pi)|\bar s u|0\rangle
=
\frac{\Delta_{K\pi}}{m_s-m_u}\,F_0(s),
\label{eq:scalarFFapp}
\end{equation}
where $m_s$ and $m_u$ denote the running quark masses evaluated at the scale $\mu \sim m_\tau$. We use $m_s(2~\text{GeV}) = 93.4^{+8.6}_{-3.4}$ MeV and $m_u(2~\text{GeV}) = 2.16^{+0.49}_{-0.26}$ MeV~\cite{Zyla:2020zbs}.

Finally, the tensor current matrix element is parameterized as~\cite{Gonzalez-Alonso:2016yxt}
\begin{equation}
\langle K(p_K)\pi(p_\pi)|\bar s \sigma_{\mu\nu} u|0\rangle
=
i\,\frac{F_T(s)}{m_K+m_\pi}
\left(
p_{K\mu}p_{\pi\nu}-p_{K\nu}p_{\pi\mu}
\right).
\label{eq:tensorFFapp}
\end{equation}
The tensor form factor $F_T(s)$ has been studied using chiral perturbation theory, dispersion relations, and lattice QCD~\cite{Bernard:2014vza,Gonzalez-Alonso:2016yxt}.

The kaon three-momentum entering Eq.~\eqref{eq:diffdist} is $|\vec p_K|=\lambda^{1/2}(s,m_K^2,m_\pi^2)/(2\sqrt{s})$, where $\lambda(a,b,c)=a^2+b^2+c^2-2(ab+bc+ca)$ is the K\"all\'en function.

\subsection{Angular Coefficients}

Retaining the dominant vector, scalar, and tensor contributions, the angular coefficients of Eq.~\eqref{eq:diffdist} can be written schematically as~\cite{Delepine:2007qg,Li:2020vru}
\begin{align}
\mathcal{A}(s) &= |1+g_V|^2 \left[ |\vec{p}_K|^2 |F_+(s)|^2 + \frac{m_\tau^2}{4s} \left|\frac{\Delta_{K\pi}}{\sqrt{s}}F_0(s)\right|^2 \right] \nonumber \\
&\quad + |g_S|^2 \frac{m_\tau^2}{4s} \left|\frac{\Delta_{K\pi}}{\sqrt{s}}F_0(s)\right|^2 \nonumber \\
&\quad + |g_T|^2 \frac{16|\vec{p}_K|^4}{s(m_K + m_\pi)^2} |F_T(s)|^2 \nonumber \\
&\quad + \text{Re}[(1+g_V)g_S^*] \frac{m_\tau^2 \Delta_{K\pi}}{s\sqrt{s}} \text{Re}[F_+(s)F_0^*(s)]
\label{eq:AcoeffApp}
\\[2mm]
\mathcal C(s) &=
-|1+g_V|^2\,|\vec p_K|^2 |F_+(s)|^2
+\mathcal C_T(s),
\label{eq:CcoeffApp}
\end{align}
where the scalar and tensor pieces additionally contribute through
\begin{align}
\mathcal A_S(s) &\propto |g_S|^2 |F_0(s)|^2,
&
\mathcal A_T(s) &\propto |g_T|^2 |F_T(s)|^2,
&
\mathcal C_T(s) &\propto |g_T|^2 |F_T(s)|^2.
\end{align}
The coefficient $\mathcal B(s)$, given in Eq.~\eqref{eq:Bcoeff} of the main text, is generated by interference between amplitudes of different Lorentz structure and is the only coefficient relevant to the forward-backward observables of Sec.~\ref{sec:eff}.

\subsection{Schematic Form of the Integrated CP Asymmetry}

When the tensor operator dominates, the integrated CP asymmetry of Eq.~\eqref{eq:ACPdef} takes the schematic form
\begin{equation}
A_{\rm CP}
\propto
\frac{
\displaystyle
\int_{(m_K+m_\pi)^2}^{m_\tau^2} ds\,
\left(1-\frac{s}{m_\tau^2}\right)^2
|\vec p_K|^3\,
\text{Im}(g_T)\,
\text{Im}\!\left[F_+(s)F_T^\ast(s)\right]
}{
\displaystyle
\int_{(m_K+m_\pi)^2}^{m_\tau^2} ds\,
\left(1-\frac{s}{m_\tau^2}\right)^2
|\vec p_K|^3\,|F_+(s)|^2
}.
\label{eq:ACPschematicApp}
\end{equation}

\section{Weak-Phase Structure: A Discrete-Symmetry Origin for the Benchmark Phases}
\label{sec:phases}

The benchmark scenario of Sec.~\ref{sec:num} sets the right-handed CKM phase and the scalar Yukawa phase to their maximal values, $\alpha_R = \phi_S = -\pi/2$. As presented, this choice is a numerical input rather than a prediction, and we now address to what extent it can be motivated rather than simply assumed.

We recall that the LRIS model already requires an additional discrete symmetry -- taken to be $\mathbb{Z}_2$ or $\mathbb{Z}_4$ in Sec.~\ref{sec:LRIS} -- to forbid unwanted large singlet mass terms and enforce the inverse-seesaw texture. If this same discrete symmetry is extended to act on the scalar sector (the bidoublet $\phi$ and the doublet $\chi_R$ of Sec.~\ref{sec:LRIS}), it constrains not only the fermionic textures but also which terms are allowed in the scalar potential, and can render the relevant CP-violating phases \emph{calculable} rather than free, along the lines of the "geometrical CP violation" mechanism first identified by Branco, Gerard, and Grimus~\cite{Branco:1984ka} (see also~\cite{Lee:1973iz}) in multi-scalar potentials with discrete symmetries.

To see this concretely, consider a $\mathbb{Z}_4$ symmetry acting on the relevant scalar bilinear $X \equiv \phi^\dagger \chi_R \sim |v_1||v_2|\, e^{i\theta}$ as $X \to iX$. The renormalizable potential built from $X$ then contains only two independent structures: $|X|^2$, which is phase-independent and therefore does not fix $\theta$, and $\mathrm{Re}[X^4] \propto \cos(4\theta)$, which is the lowest $\mathbb{Z}_4$-invariant term sensitive to the phase. Minimizing $V(\theta) = -\kappa\cos(4\theta)$ with respect to $\theta$ gives extrema at $\theta = n\pi/4$ ($n=0,1,2,3$), with $\theta = 0 \pmod{\pi/2}$ a minimum for $\kappa>0$ (the CP-conserving solution) and $\theta = \pi/4 \pmod{\pi/2}$ a minimum for $\kappa<0$ (a CP-violating solution with a phase \emph{quantized} by the symmetry, rather than continuously tunable). The sign of $\kappa$ is a discrete, technically natural choice -- not a fine-tuned continuous parameter -- and the resulting phase is either exactly zero or a fixed rational multiple of $\pi$, never a generic $\mathcal{O}(0.1)$ value. Depending on how many powers of the relevant bidoublet combination enter the specific Yukawa coupling responsible for $\alpha_R$ or $\phi_S$, this quantized potential-level phase can map onto a maximal ($\pi/2$) phase in the physical Yukawa coupling.

We emphasize that this argument is a plausibility mechanism illustrating why phases near $\pi/2$ are technically natural in this class of models, not a complete derivation for the full LRIS scalar potential (which involves additional terms among $\phi$, $\chi_R$, and the SM-like doublet, and is left for future work). We also are not aware of a direct experimental constraint (e.g.\ from electron or neutron EDMs) that currently bounds $\alpha_R$ or $\phi_S$ individually within this model; a dedicated EDM analysis, which would require computing the relevant one- and two-loop diagrams induced by the charged and neutral scalar sector, is beyond the scope of the present work.

\begin{figure}[htbp]
\centering
\includegraphics[width=0.85\linewidth]{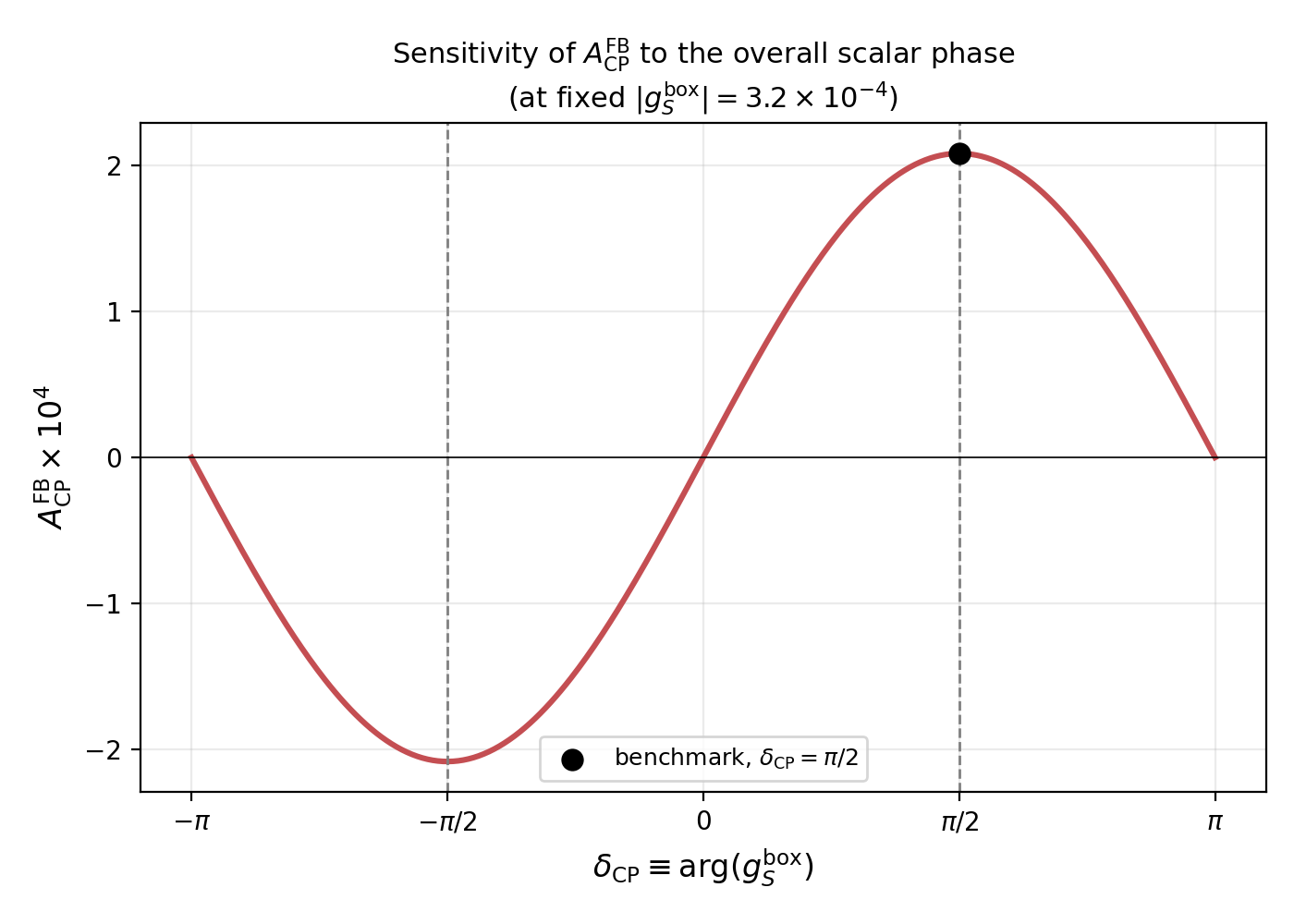}
\caption{Sensitivity of the integrated forward-backward CP asymmetry $A_{\rm CP}^{\rm FB}$ to the overall CP-violating phase $\delta_{\rm CP} \equiv \arg(g_S^{\rm box})$, at fixed $|g_S^{\rm box}| = 3.2\times 10^{-4}$ [Eq.~\eqref{eq:gSnumeric}]. The dependence follows $\sin\delta_{\rm CP}$ exactly [Eq.~\eqref{eq:AFBsensitivity}]. The benchmark point $\delta_{\rm CP}=\pi/2$ (black dot) corresponds to the maximal, symmetry-motivated phase discussed in Sec.~\ref{sec:phases}.}
\label{fig:phasesens}
\end{figure}

Given the significance of the overall phase for the observability of $A_{\rm CP}^{\rm FB}(s)$, it is useful to examine its dependence explicitly. Defining $\delta_{\rm CP} \equiv \arg(g_S^{\rm box})$, Eq.~\eqref{eq:AFBCP} shows that, at fixed $|g_S^{\rm box}|$ and for $g_V \approx 0$, the CP-odd interference term scales as $\mathrm{Im}[(1+g_V)g_S^{\rm box\,*}] = |g_S^{\rm box}|\sin\delta_{\rm CP}$, so that the integrated forward-backward asymmetry itself follows
\begin{equation}
A_{\rm CP}^{\rm FB} (\delta_{\rm CP})\;=\; A_{\rm CP}^{\rm FB,\,LRIS}\Big|_{\delta_{\rm CP}=\pi/2} \times \sin\delta_{\rm CP}\,,
\label{eq:AFBsensitivity}
\end{equation}
an exact consequence of Eq.~\eqref{eq:AFBCP} rather than an approximation. Figure~\ref{fig:phasesens} shows this dependence explicitly. The benchmark value of $2.08\times 10^{-4}$ [Eq.~\eqref{eq:AFBCPnumeric}] corresponds to the maximal phase $\delta_{\rm CP}=\pi/2$ singled out by the discrete-symmetry argument above; away from this point the observable degrades smoothly, remaining within reach of Belle~II down to $|\sin\delta_{\rm CP}| \gtrsim 0.5$, and vanishes, as expected, for a CP-conserving phase $\delta_{\rm CP} = 0, \pi$.

%

\end{document}